\documentclass[a4paper, 12pt]{article}
\pdfoutput=1

\usepackage{latexsym,amsmath,amsfonts,amssymb}
\usepackage{tikz}
\usetikzlibrary{decorations.pathmorphing,cd,decorations.markings,calc}
\usepackage{mathrsfs}
\usepackage[american]{babel}
\usepackage{graphicx}
\usepackage{bbm}
\usepackage{cite}
\usepackage{tcolorbox}
\usepackage{cancel}
\usepackage{appendix}

\usepackage{hyperref}
\usepackage{cleveref}

\definecolor{pansypurple}{rgb}{0.47, 0.09, 0.29}
\definecolor{patriarch}{rgb}{0.5, 0.0, 0.5}
\definecolor{carmine}{rgb}{0.59, 0.0, 0.09}
\definecolor{blueflag}{rgb}{0.2, 0.2, 0.6}
\definecolor{violet(ryb)}{rgb}{0.53, 0.0, 0.69}
\definecolor{operamauve}{rgb}{0.72, 0.52, 0.65}
\definecolor{olive}{rgb}{0.42, 0.56, 0.14}
\definecolor{mulberry}{rgb}{0.77, 0.29, 0.55}
\definecolor{prettyyellow}{rgb}{0.91, 0.84, 0.42}
\definecolor{electricviolet}{rgb}{0.56, 0.0, 1.0}
\definecolor{persianrose}{rgb}{1.0, 0.16, 0.64}

\renewcommand{\baselinestretch}{1.2}
\newcommand{\bt}{\begin{tikzcd}}
\newcommand{\et}{\end{tikzcd}}

\numberwithin{equation}{section}

\newcommand{\be}{\begin{equation}} \newcommand{\ee}{\end{equation}}
\newcommand{\bea}{\begin{equation} \begin{aligned}} \newcommand{\eea}{\end{aligned} \end{equation}}

\newcommand{\cD}{\mathcal{D}}

\newcommand{\cM}{\mathcal{M}}

\newcommand{\cN}{\mathcal{N}}

\newcommand{\bQ}{\mathbb{Q}}
\newcommand{\bR}{\mathbb{R}}

\newcommand{\bZ}{\mathbb{Z}}

\newcommand{\unit}{\mathbbm{1}}

\newcommand{\one}{^{(1)}}

\def\repa{\raise4pt\hbox{$\square$}\mkern-14mu\raise-4pt\hbox{$\square$}}
\def\repab{\overline{\raise4pt\hbox{$\square$}\mkern-14mu\raise-4pt\hbox{$\square$}\mkern-1mu}}

\begin{document}
\thispagestyle{empty}
	\fontsize{12pt}{20pt}	
	\vspace{13mm}  

 \color{black}
	\begin{center}
		{\huge  

    4d Maxwell on the Edge: Global Aspects of Boundary Conditions and Duality
    
  }

\vskip 20pt

		{\large Adrien Arbalestrier,  Riccardo Argurio, Giovanni Galati, 
        and Elise Paznokas}	
				\bigskip
				
				{\it
						 Physique Th\'eorique et Math\'ematique and International Solvay Institutes\\
Universit\'e Libre de Bruxelles, C.P. 231, 1050 Brussels, Belgium  \\
					
					}
		
	\end{center}

\begin{abstract}

We revisit Maxwell theory in 4d with a boundary, with particular attention to the global properties of the boundary conditions, both in the free (topological) and interacting (conformal) cases. We analyze the fate of Wilson-’t Hooft lines, identifying the subset that is trivialized on the boundary and the ones that become topological, thus generating a boundary 1-form symmetry. We further study how the boundary conditions are mapped to each other by 3d topological interfaces implementing bulk dualities and rescalings of the coupling. Together, these interfaces generate an $SL(2,\bQ)$ action on the bulk complexified coupling $\tau$, and they generalize the usual $SL(2,\bZ)$ action on 3d CFTs by including both topological and non-topological manipulations within a unified framework. We then show how to recover our results in a streamlined way from a SymTFT picture in 5d with corners. Finally, we comment on the possible inclusion of non-compact 3d edge modes.

\end{abstract}

\newpage
\pagenumbering{arabic}
\setcounter{page}{1}
\setcounter{footnote}{0}
\renewcommand{\thefootnote}{\arabic{footnote}}

{\renewcommand{\baselinestretch}{.88} \parskip=0pt
	\setcounter{tocdepth}{2}
	\tableofcontents}


\section{Introduction}
Quantum Field Theories defined on spaces with boundaries are of great interest for several reasons. In the study of classical or quantum materials, boundaries are unavoidable in experimental setups, making it essential to develop methods that properly account for boundary contributions. From a theoretical point of view, there is no general systematic procedure to classify the full space of consistent boundary conditions in a given QFT, with a few notable exceptions including two-dimensional rational CFTs \cite{Cardy:2004hm} or certain higher-dimensional supersymmetric theories (see e.g. \cite{Gaiotto:2008sa,Gaiotto:2008ak}). Even in relatively simple cases, such as free theories, this classification turns out to be surprisingly difficult and often reveals unexpected features \cite{Giombi:2019enr, Behan:2020nsf, DiPietro:2020fya, Behan:2021tcn, DiPietro:2023gzi}. For these reasons, the study of boundary conditions frequently provides valuable insights into non-perturbative properties of the QFT under consideration. Finally, boundary conditions of QFTs in asymptotically AdS spacetimes play a key role in probing the AdS/CFT correspondence from a bottom-up perspective and, more recently, have emerged as a useful tool for regularizing flat-space observables \cite{Polchinski:1999ry,Fitzpatrick:2010zm,Penedones:2010ue,Paulos:2016fap} and for probing interesting features such as mass gap and confinement \cite{Aharony:2012jf,Copetti:2023sya,Ciccone:2024guw}.

When dealing with gauge theories, it is by now well understood that, beyond the spectrum of local operators and their associated observables, the global properties of the theory, also referred to as \emph{global structure}, arise through the study of their extended operators \cite{Aharony:2013hda} and the corresponding generalized symmetries that measure their charges \cite{Gaiotto:2014kfa}. The goal of this paper is to analyze how these properties manifest when the theory is considered in spaces with boundaries.\footnote{The interplay between boundaries, defects, and global symmetries has recently received renewed attention, leading to several interesting insights, see e.g. \cite{Padayasi:2021sik,Drukker:2022pxk,Aharony:2023amq,Copetti:2024onh,Antinucci:2024izg,Barkeshli:2025cjs,Antinucci:2025uvj,Choi:2025ebk,Copetti:2025sym,Komargodski:2025jbu}.} In particular, while they constrain local observables, boundary conditions also encode non-trivial global structures that control the behavior of extended operators near the boundary. They specify which operators become trivial, which turn into topological ones—thus defining boundary higher-form symmetries—and which instead carry non-trivial charges under bulk and boundary symmetries. From this perspective, four-dimensional Maxwell theory provides a paradigmatic example, thanks to its rich structure of line and surface operators, their evolution along the conformal manifold parametrized by the complexified coupling $\tau$, and their interplay with electric-magnetic duality. Our main tool is the description of boundary conditions in terms of three-dimensional Topological QFTs, appropriately coupled to the bulk photon, as first introduced in \cite{Kapustin:2009av} (see also \cite{Choi:2023xjw,Cordova:2023ent,Apruzzi:2024htg} for recent discussions). When these are the only boundary modes, the corresponding boundary conditions define an exactly solvable class of boundary theories, which we refer to as free boundary conditions, following the notation of \cite{DiPietro:2019hqe}. Another interesting class of boundary conditions are those which preserve the maximal subgroup of the bulk conformal symmetry, dubbed conformal boundary conditions. These can be described by the addition of $U(1)$ symmetric 3d CFTs coupled, together with the TQFT, to the bulk. 

The topological edge mode description provides a well-defined and useful setup to analyze in detail both the local and global properties of the boundary. In the case of the free boundary conditions, we find that inequivalent ones are characterized by three integers $(P,Q,\tilde{r})$ with $\gcd(P,Q)=1$. While $P$ and $Q$ govern the fate of the local observables on the boundary, giving the condition
\begin{equation}
P\left(\frac{i}{e^2} *F - \frac{\theta}{2\pi} F\right) + QF = 0 \,,
\end{equation}
the integer $\tilde{r}$ determines the allowed endable line operators, setting
\begin{equation}\label{eq: lines}
    (W_QH_P(\gamma))^{m} = 1\quad,\quad \forall  m\in \tilde{r} \, \bZ\,,
\end{equation}
where $W_Q$ and $H_P$ are Wilson and 't Hooft lines with charges $Q$ and $P$, respectively. It also determines the $\bZ_{\tilde r}$ 1-form boundary symmetry generated by the bulk line operators in \eqref{eq: lines} with $m \in \bZ$ that become topological when restricted to lie on the boundary. When dealing with conformal boundary conditions, the topological couplings select which subset of the boundary charged operators are coupled to the bulk, thus implementing the precise map between bulk and boundary global symmetries.

The bulk $U(1)_e \times U(1)_m$ 1-form symmetry enjoyed by the free photon \cite{Gaiotto:2014kfa} provides a natural action on the space of boundary conditions. The latter can be implemented by fusing a generic topological interface which separates two Maxwell theories, which is obtained by gauging a (finite) anomaly free global symmetry on half space, onto the boundary. We find that these interfaces map the parameters $(P,Q)$ to an  $SL(2,\mathbb{Z})$ transformed doublet, while also acting on the parameter $\tilde{r}$ in a tractable way. In the case of conformal boundary conditions, such maps give rise to a refined version of the $SL(2,\mathbb{Z})$ transformations on 3d CFTs described in \cite{Witten:2003ya}, where both non-topological and topological manipulations are taken into account. We provide a detailed analysis of this map and discuss its implications in the weakly coupled limit, where the gauge coupling $e$ is set to zero and the bulk decouples from the boundary dynamics. In this case, we find several local three-dimensional CFTs, all connected to a parent theory through an intricate sequence of topological and non-topological manipulations. In more detail, the set of theories that are mapped to each other is the one of 3d CFTs with a $U(1)$ 0-form symmetry and a $\bZ_n$ 1-form symmetry. The action on this set is by (non-topologically) gauging the $U(1)$, by adding a Chern-Simons term for its background field (respectively, the $S$ and $T$ transformations of $SL(2,\bZ)$), and by (topologically) gauging a finite subgroup of either $U(1)$ or $\bZ_n$. Seen from the bulk, these two kinds of actions map to interfaces that act generically as $SL(2,\bQ)$ on the coupling $\tau$, but that require a finer definition to encode their potential non-invertibility.

The setup of 4d Maxwell theory also provides a natural playground for exploring the holographic interpretation of these results. This is realized through the Symmetry TFT (SymTFT) formulation, namely a 5d TQFT defined on a slab manifold, which serves as the bulk dual once suitable boundary conditions are imposed on the two sides of the slab \cite{Kong:2014qka,Pulmann:2019vrw,Gaiotto:2020iye,Apruzzi:2021nmk,Freed:2022qnc}. In line with \cite{Bhardwaj:2024igy,Choi:2024tri,GarciaEtxebarria:2024jfv}, when the four-dimensional theory is placed on a manifold with boundaries,\footnote{See also \cite{Bonetti:2025dvm}, which analyzes a SymTFT setup with corners to investigate spontaneous symmetry breaking, and \cite{Cvetic:2024dzu}, which explores QFTs localized at the corner of a holographic bulk geometry.} the corresponding five-dimensional geometry can be modified by introducing corners, as illustrated on the left of Fig.~\ref{edges symtft}. A more refined construction is obtained by introducing a new topological boundary interpolating between the physical and the topological sides of the slab, see the right of Fig.~\ref{edges symtft}. In this way, the topological couplings of the boundary conditions are nicely captured by the possible choices of the new topological boundary conditions, denoted by $\mathcal{B}_{4d' ,\mathrm{top.}}$, together with the associated 3d topological corner. This framework also offers a natural interpretation of the action of topological interfaces on boundary conditions, which can be recovered via an oblique compactification of the setup, a procedure which we analyze in detail in this work.

The paper is organized as follows. In Section \ref{section 2}, we analyze free boundary conditions of Maxwell theory, where only topological edge modes are present. We describe how the TQFT determines the fate of both local and extended operators on the boundary, and identify the physical data that can be extracted from the topological couplings appearing in the action. In Section~\ref{sec: 3}, we explicitly construct the topological interfaces implementing the bulk $SL(2,\mathbb{Z})$ duality, together with those associated with rescaling of the coupling $\tau$. Taken together, these exhaust the set of topological interfaces of Maxwell theory. We then determine how such interfaces act on a given free boundary condition once they are fused onto the boundary. In Section \ref{interacting bc}, we study conformal boundary conditions, where non-trivial 3d CFTs are coupled to the bulk photon. We show how the topological action modifies the bulk to boundary coupling, and analyze the local 3d theory emerging in the weak coupling limit. In this context, we find a natural generalization of the $SL(2,\mathbb{Z})$ action on 3d CFTs. In Section \ref{setup}, we show how these results can be recovered within the SymTFT framework, where the topological properties of the boundary are fully encoded in the bulk setup. Finally, in Section \ref{sec: non compact b.c.}, we discuss possible generalizations involving non-compact edge modes on the boundary, highlighting both their implications and the potential pathologies they introduce.

\section{Boundary Conditions of Free Maxwell Theory} \label{section 2}
Let us consider an Euclidean 4-dimensional Maxwell theory, described by the action
\begin{equation}\label{eq:action4dMaxwell}
    S_{4d} = \int_{X_4} \left(\frac{1}{4\pi e^2} F \wedge * F + \frac{i\theta}{8\pi^2} F \wedge F \right)
\end{equation}
on a manifold $X_4$ with boundaries.\footnote{Let us emphasize that, in Euclidean signature, the Hodge star satisfies $*^2 = 1$ when acting on 2-forms.} Here $F$ is the curvature of a $U(1)$ connection $A$
and $\tau := \frac{i}{e^2}+\frac{\theta}{2\pi}$ parametrizes a family of (free) conformal theories. As usual, whenever we have boundaries we should impose boundary conditions for the fields. Such conditions cannot be arbitrary since they need to be compatible with the variational principle. These are usually understood as delta-functions inserted inside the path-integral, namely 
\begin{equation}
    Z = \int DA e^{-S[A]}\delta (\text{b.c.})\,.
\end{equation}
However, it is sometimes useful to modify this presentation by adding boundary terms to the action, possibly introducing new degrees of freedom—edge modes.

\subsection{Free boundary conditions}
The space of such boundary conditions is infinite, and it essentially corresponds to the class of 3d QFTs with a $U(1)$ global symmetry \cite{DiPietro:2019hqe}. However, one can focus on a particularly simple and well-behaved subset, where the theory remains quadratic and boundary conditions are imposed solely on the bulk fields without introducing additional propagating boundary degrees of freedom. We will refer to these as free boundary conditions.  Famous examples of these conditions are Dirichlet and Neumann boundary conditions, namely
\begin{equation}\label{eq:Dir/Neum}
    \begin{split}
   & \text{Dirichlet: } F|_{\partial X_4} = 0\\
   & \text{Neumann: } \left.\left(\frac{i}{e^2}* F - \frac{\theta}{2\pi} F \right)\right|_{\partial X_4}= 0\,,
   \end{split}
\end{equation}
but we will show that there are more possibilities and that \eqref{eq:Dir/Neum} is not the precise way of imposing these boundary conditions. 

Free boundary conditions can be written as 3d topological field theories coupled to the bulk \cite{Kapustin:2009av, Cordova:2023ent}. For example, Dirichlet boundary conditions can be written as
\begin{equation}\label{eq: Dir ac}
    S_{3d}^{D}= \frac{i}{2\pi}\int_{\partial X_4} \Phi \wedge dA \,,
\end{equation}
while Neumann is
\begin{equation}\label{eq: Neum ac}
    S_{3d}^{N}=\frac{i}{2\pi}\int_{\partial X_4} \Phi_1 \wedge d  \Phi_2 - \Phi_2 \wedge dA\,.
\end{equation}
Here $\Phi,\Phi_{1}$ and $\Phi_2$ are $U(1)$ gauge fields defined on the boundary $\partial X_4$. Indeed, by integrating them out, one gets
\begin{align}
    & \text{Dirichlet: }dA = 0\ , \\
    & \text{Neumann: } dA = d\Phi_1\,.
\end{align}
Moreover, the boundary equations of motion for $A$ imply
\begin{equation}\label{eq: electric cond}
    \begin{split}
    & \text{Dirichlet: } \frac{i}{e^2} * F - \frac{\theta}{2\pi} F = d\Phi \,,\\
    & \text{Neumann: } \frac{i}{e^2} * F - \frac{\theta}{2\pi} F =- d\Phi_2 = 0\,.
    \end{split}
\end{equation}
In fact, an equivalent way to encode Neumann boundary conditions is to have no edge modes at all, which indeed follows by integrating out $\Phi_2$ and then $\Phi_1$. One can alternatively consider the shift $\Phi_1\to \Phi_1+A$, which decouples the edge modes from the bulk. It straightforwardly yields the same boundary equations of motion as above.

Let us finally note that the bulk theory has a $U(1)_e^{(1)} \times U(1)_m^{(1)}$ 1-form symmetry, generated by\footnote{We choose reality conventions on the currents so that the charges $Q=\int * J$ are real. } 
\begin{equation}\label{JeJm}
    * J_e = -\frac{i}{2\pi e^2}* F + \frac{\theta}{4\pi^2} F\equiv \frac{1}{2\pi} \tilde{F}  \quad,\quad * J_m = \frac{1}{2\pi} F \,,
\end{equation}
which act respectively on Wilson and 't Hooft lines defined in terms of the gauge field $A$ and its dual $\tilde{A}$ as
\begin{equation}
    W_n[\gamma] = \exp{\left(i n \int_{\gamma} A\right)} \quad, \quad H_m[\gamma]=\exp{\left(-i m \int_{\gamma} \tilde{A}\right)}\, ,
\end{equation}
where $n,m\in\bZ$ are respectively the charges of these operators under the electric and magnetic symmetries.\footnote{The minus sign in the definition of the 't Hooft line $H_m[\gamma]$ in terms of the dual field $\tilde{A}$ follows from using the same conventions as in \cite{Niro:2022ctq}.} The Dirichlet (resp.~Neumann) boundary conditions defined in \eqref{eq:Dir/Neum} preserves the $U(1)$ global symmetry generated by $J_e$ (resp.~$J_m$), yielding a $U(1)$ 0-form symmetry on the boundary. 

\subsubsection{Global properties of the boundary conditions} \label{global section}
The crucial subtlety that we want to focus on is that the conditions \eqref{eq:Dir/Neum} do not fully determine the boundary conditions. While they fix the field strength, they leave the holonomy of the gauge field undetermined—an additional gauge-invariant piece of data associated with the principal bundle. However, this piece of data is completely constrained by the actions \eqref{eq: Dir ac} and \eqref{eq: Neum ac}; by summing over the flux sectors $\int d\Phi_i\in 2\pi \bZ$, we get the conditions
\begin{align}
    & \text{Dirichlet: } \int A \in 2\pi \bZ \,,\\
    & \text{Neumann: }  \int A = \int \Phi_1\,.
\end{align}
We conclude that for Dirichlet boundary conditions, all the Wilson lines become trivial on the boundary. In contrast, in the Neumann case, they are not.

A similar discussion can be done for the 't Hooft lines of the theory. The constraints of \cref{eq: electric cond} imply that the 't Hooft lines on the boundary can be written as
\begin{align}
     & \text{Dirichlet: } \quad  H_q[\gamma] \equiv \exp{\left(i q\int \Phi\right)}\,,\\
    & \text{Neumann: } \quad H_q[\gamma] \equiv \exp{\left(-i q\int \Phi_2\right)} = 1 \quad \forall q\,.
\end{align}

Crucially, the conditions on the lines can, in fact, be relaxed while preserving the local conditions on the field strength. For example, we can consider the modified actions
\begin{align}
    & \text{Dirichlet: } \quad  S_{3d}^{D,\tilde r}= \frac{i\tilde r}{2\pi}\int_{\partial X_4} \Phi \wedge dA \,,\\
    & \text{Neumann: } \quad   S_{3d}^{N, \tilde r}= \frac{i}{2\pi}\int_{\partial X_4}  \tilde r \Phi_1 \wedge d \Phi_2-\Phi_2 \wedge dA \,.
\end{align}
where $\tilde r \in \bZ$ in order to preserve gauge invariance. The e.o.m. still imply \eqref{eq:Dir/Neum}. However, now, the sum over the quantized fluxes of the edge modes gives the modified constraints
\begin{align}
   & \text{Dirichlet: } \int A \in \frac{2\pi}{\tilde r} \bZ\ , \\
    &\text{Neumann: } \int \Phi_2 \equiv \int \tilde{A} \in \frac{2\pi}{\tilde r} \bZ\,,
\end{align}
namely only those Wilson ('t Hooft) lines with charge $q \in \tilde r\bZ $ are trivialized. Moreover, since $dA =0$ ($d \tilde{A} =0$) at the boundary, the non-trivial ones become topological there, generating a $\bZ_{\tilde r}^{(1)}$ boundary 1-form symmetry. Physically, this modified Dirichlet (Neumann) boundary condition is obtained from the standard one by gauging the $\bZ_{\tilde r}$ subgroup of the boundary 0-form global symmetry.

\subsubsection{Generic free boundary conditions} \label{section 2.1.2}
As emphasized earlier, Dirichlet and Neumann boundary conditions do not generate the full set of free boundary conditions for Maxwell theory. To describe a general free boundary condition, one can consider the most general 3d Abelian topological field theory that can be coupled to the bulk. 
This leads to the following action \cite{Kapustin:2009av}:
\begin{equation} \label{3d phys}
    S_{3d} = \frac{i}{2\pi} \int_{M_3} \left( \frac{p}{2} A \wedge dA + A \wedge (v^T\, d\Phi) + \frac{1}{2} \Phi^T \wedge (k\,d\Phi) \right),
\end{equation}
where $p\in\mathbb{Z}$, $\Phi=(\Phi_1,\cdots, \Phi_n)^T$ is a vector of 1-form $U(1)$ gauge fields living on the 3d boundary, $k_{j\ell}$ ($j,\ell=1,\dots,n$) is a symmetric integral matrix (ensuring proper quantization under large gauge transformations), and $v_\ell$ ($\ell=1,\dots,n$) is a vector of integers. 

These boundary conditions are therefore characterized by the data \((p, v, k)\). However, different choices of \((p, v, k)\) may lead to physically equivalent boundary conditions at the quantum level. For instance, we have the freedom to redefine the fields—without violating the \(U(1)\) quantization conditions—such that the boundary action is rewritten with transformed coefficients:
\begin{equation}
    k' = S k S^T\,, \qquad v' = S v\,,
\end{equation}
by considering the redefinition $\Phi=S^T\Phi'$, with \(S \in SL(n,\mathbb{Z})\). Moreover, we can rewrite the Chern-Simons coupling for $A$ by adding an extra edge mode as
\begin{equation}
    \int \frac{ip}{4\pi} A \wedge dA = \int \frac{i p }{2\pi} A \wedge d \Phi_{n+1} - \frac{ip}{4\pi} \Phi_{n+1} \wedge d\Phi_{n+1}
\end{equation}
such that 
\begin{equation}
    (p,v,k) \simeq (0,v',k')\ , \quad {v'} = \begin{pmatrix}
       v  \\ p
    \end{pmatrix}\ , \quad k' = \begin{pmatrix}
        k & 0 \\ 0 & -p \\
    \end{pmatrix} \,.
\end{equation}
Notice that, using these redundancies, one is able to express a generic boundary condition as\footnote{Using the $SL(n,\bZ)$ subgroup of $SL(n+1,\bZ)$ leaving $v$ invariant, one can further restrict $k$ to be a symmetric tridiagonal matrix, i.e. $k_{ij}\neq 0$ only if $|i-j|\leq 1$.}
\begin{equation}
    (p=0,\,v = (v_1,0,\cdots,0)^T , \,k)\,.
\end{equation}
As before, the equations of motion for \(\Phi_\ell\) impose local constraints on the bulk field strength. Assuming that the matrix \(k\) is non-degenerate,\footnote{The case of a degenerate $k$, which is related to Dirichlet-like boundary conditions, will be treated at the end of this subsection.} we obtain:
\begin{equation}
    \frac{i}{e^2} *F - \frac{\theta}{2\pi} F + \left( v^T k^{-1} v-p\right) F = 0\,.
\end{equation}
The combination \( v^T k^{-1} v-p\) is intrinsic to the boundary condition and invariant under the redundancies discussed before. To ensure gauge invariance, we considered integer components for $p$, $v$ and $k$. This implies that this combination is a rational number. Therefore, the general form of the boundary condition becomes:
\begin{equation} \label{27}
    P\left(\frac{i}{e^2} *F - \frac{\theta}{2\pi} F\right) + QF = 0 \quad,\quad \frac{Q}{P} :=  v^T k^{-1} v-p\quad , \quad \gcd(P,Q)=1\,,
\end{equation}
in line with the discussion in \cite{DiPietro:2019hqe}. This boundary condition implies that, on the boundary, the dyonic lines $(W_QH_P)^q$ become topological operators for any $q\in \bZ$.  Note that these are exactly the lines that are uncharged under the symmetry that is trivialized on the boundary, which is the one generated by $QJ_m-PJ_e$. This is consistent with the fact that some of these lines actually turn out to be trivial on the boundary, as we now show. 

In addition to this local condition, the boundary data also determines the fate of the bulk Wilson and ’t Hooft lines when defined on the boundary, thereby fully specifying the boundary condition. In order to determine the global structure it is convenient to consider the equivalent presentation of the family of boundary conditions as
\begin{equation}\label{3d phys equiv}
    S_{3d} = \frac{i}{2\pi} \int_{M_3}  A \wedge  (v^T\, d\Phi) + \frac{1}{2}  \Phi^T \wedge (k\, d\Phi) \,.
\end{equation}
Summing over the holonomies of $\Phi_\ell$ gives the $n+1$ following constraints: 
\begin{equation}\label{global constraints Kapustin}
    \int v\, A+k \Phi\in 2\pi \bZ^{n+1}.
\end{equation}
Since $k$ is a matrix of integers and is assumed to be non-degenerate, $k^{-1}$ is a matrix of rational numbers. In particular, the components of the vector $k^{-1}v$ are rational numbers and there exists an integer $R$ such that $Rk^{-1}v$ is a vector of integers.\footnote{Note that we can always choose $R=\det(k)$, but in general there can be smaller choices of $R$ that satisfy the same condition. At the moment, it is not necessary to impose $R$ to be the smallest such integer.} We now consider the following linear combination with integer coefficients of the constraints \eqref{global constraints Kapustin}:
\begin{equation}
    R v^T k^{-1} \int v\, A+k\Phi\in g 2\pi \bZ
\end{equation}
where $g$ is the gcd of the components of the vector $Rk^{-1}v$.\footnote{Any linear combination of integers $n_i$ with integer coefficients is proportional to $\gcd(\{n_i\})$.}
This expression can be written as follows
\begin{equation}
    R \left(\int v^Tk^{-1}v\, A+v^T \Phi\right)\in g 2\pi \bZ \,.
\end{equation}
Since $v^Tk^{-1}v=\frac{Q}{P}$, and $Rv^Tk^{-1}v\in \bZ$ we conclude that $R$ must be a multiple of $P$. Using the notation $R=rP$, we get\footnote{Since we did not impose $R$ to be the smallest integer such that $Rk^{-1}v\in\bZ^n$, it is possible to have $\gcd(r,g)\neq 1$.}
\begin{equation}
    r \left(\int Q A+P v^T \Phi\right)\in g 2\pi \bZ\,.
\end{equation}
We now consider the equation of motion obtained by varying with respect to $A$:
\begin{equation}
    \frac{i}{e^2}*F-\frac{\theta}{2\pi}F=-\tilde{F} = v^T d\Phi\,.
\end{equation}
We conclude that, on the 3d boundary, 't Hooft lines of the 4d bulk become 
\begin{equation}\label{eq: thooft line}
    H_m(\gamma)=\exp\left(im \int_{\gamma} v^T \Phi\right) \,,
\end{equation}
and the combination $-v^T \Phi$ can be interpreted as the dual field $\tilde{A}$. The global boundary condition can then be written as
\begin{equation}
    r \left(\int Q A-P \tilde{A}\right)\in g 2\pi \bZ\,.
\end{equation}
Therefore, the dyonic lines trivialized on the 3d boundary are
\begin{equation} \label{34}
    (W_QH_P(\gamma))^{\tilde{r} \,m}\quad,\quad \tilde{r}:= \frac{r}{\gcd(r,g)} \quad ,\quad \forall  m\in \bZ\,.
\end{equation}
Note that while we have freedom to choose different values of $R$, which will change $r$ and $g$, the resulting $\tilde{r}$ is invariant under this choice. We conclude that the dyonic line of the form $W_QH_P$ defines a $\bZ_{\tilde{r}}$ 1-form symmetry of the boundary. Line operators of charge $q$ under this symmetry are 
\begin{equation}\label{charged operator Zr/gcd}
    \exp\left(i \int_{\gamma}\textbf{q}^T \Phi\right)
\end{equation}
where $\textbf{q}$ is a vector of integers satisfying $\textbf{q}^T (Rk^{-1}v)=\gcd(r,g)\ q$. Note that when $\textbf{q}=v$, this operator corresponds to an 't Hooft line and its charge is $q=\tilde{r}\, Q\in \tilde{r}\,\bZ$. Therefore, 't Hooft lines are not charged under this $\bZ_{\tilde r}$ symmetry. As for Wilson lines, one can see from the equations of motion that they also have a trivial charge. This implies that the potential 't Hooft anomaly of this symmetry actually vanishes.

Let us finally consider the case in which $k$ is degenerate. In this instance, it possesses an eigenvector $v_0$ such that $kv_0=0$. Since $k$ is a matrix of integers, we can choose a vector $v_0$ with integer coefficients. Moreover, it is always possible to normalize this vector to have $\gcd(\{v_0^\ell\})=1$. Using a linear combination of equations \eqref{global constraints Kapustin}, we obtain
\begin{equation}
    v_0^T\left(\int v\, A+k \Phi\right)\in 2\pi \bZ\ .
\end{equation}
We conclude that
\begin{equation}
    v_0^Tv\int  A\in 2\pi \bZ\ ,
\end{equation}
thus obtaining a $\bZ_{v_0^Tv}$ boundary 1-form symmetry. Note that $v_0^Tv\in \bZ$ is invariant under $SL(n,\bZ)$ transformations since we must have $v_0\to (S^{T})^{-1}v_0$ in order to preserve $kv_0=0$.\footnote{If $v^T_0v=0$ then we obtain a trivial condition, consistently with the fact that the field along this zero eigenvector drops out from the action. One can then remove this redundancy and redefine $v$ and $k$ with fewer entries.} This case thus corresponds to a modified Dirichlet condition, with
\begin{equation}
    Q=1\,, \quad P=0\,, \quad \tilde r=v^T_0v\, .
\end{equation}

We conclude this Section by emphasizing that there can exist inequivalent boundary topological couplings (i.e.~not equivalent under the identifications described above) which lead to the same values of $P$, $Q$, and $\tilde r$. These boundary TQFTs therefore produce the same boundary conditions for the bulk fields, but they can differ in certain topological observables associated with edge modes.\footnote{For instance, one may have different numbers of boundary lines constructed from the edge modes $\Phi_i$, with distinct linking pairings.}

\section{Topological Interfaces and Boundary Conditions}\label{sec: 3}

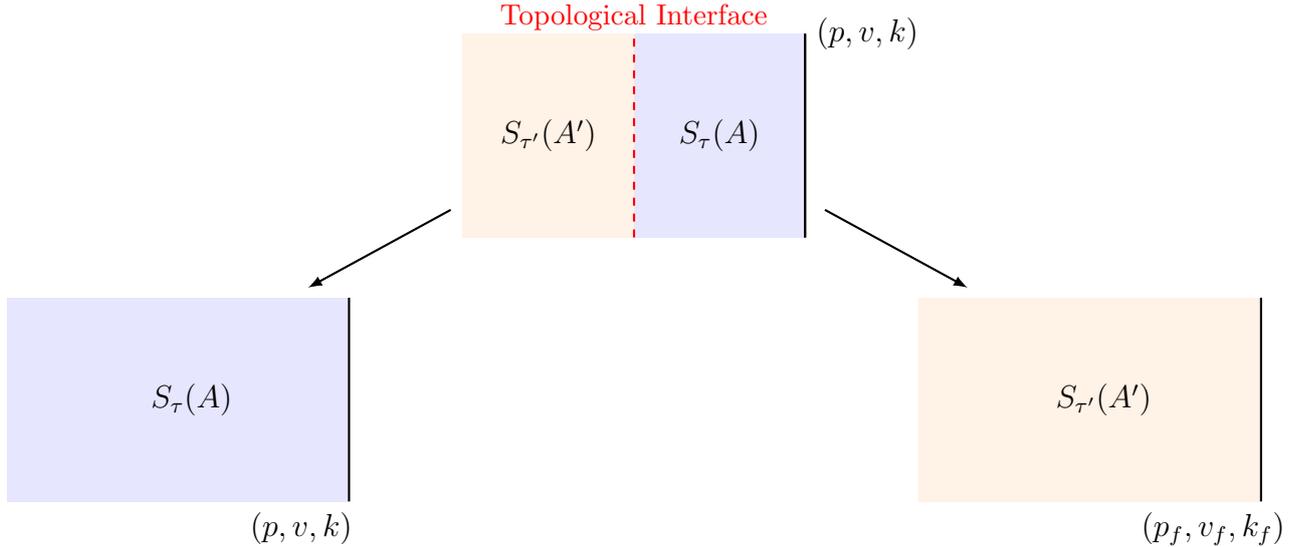
\begin{figure}[t]
    \centering
\begin{tikzpicture}[
    arrow_style/.style={->, thick, >=latex, shorten >=2pt, shorten <=2pt}, 
    arc_arrow_style/.style={->, thick, >=latex, shorten >=2pt, shorten <=2pt}
]

\begin{scope}[xshift=0cm, yshift=0cm, scale=0.9]
    \fill[blue!10] (-3,0) rectangle (2,3);
    \draw[thick] (2,0) -- (2,3); 
    \node at (-0.3, 1.5) {$S_\tau(A)$}; 
    \node at (1.3, -0.) [anchor=north] {$(p,v,k)$};
\end{scope}

\begin{scope}[xshift=6cm, yshift=3.5cm, scale=0.9]
    \fill[orange!10] (-3,0) rectangle (-0.5,3);
    \fill[blue!10] (-0.5,0) rectangle (2,3);
    \draw[thick] (2,0) -- (2,3); 
    \draw[dashed, thick, red] (-0.5,0) -- (-0.5,3) node[anchor=south, red, yshift=-1mm] {\small Topological Interface};
    \node at (-1.75, 1.5) {$S_{\tau'}(A')$};
    \node at (0.75, 1.5) {$S_\tau(A)$};
  \draw[thick] (2,0) -- (2,3) node[anchor=west] {$(p,v,k)$};
\end{scope}

\begin{scope}[xshift=12cm, yshift=0cm, scale=0.9]
    \fill[orange!10] (-3,0) rectangle (2,3);
    \draw[thick] (2,0) -- (2,3); 
    \node at (-0.3, 1.5) {$S_{\tau'}(A')$};
    \node at (1.3, -0.) [anchor=north] {$(p_f,v_f,k_f)$}; 
\end{scope}

\draw[arc_arrow_style, bend right=20] (3.2, 3.9) -- (1.2, 2.8);
\draw[arc_arrow_style, bend left=20] (8, 3.9) -- (10, 2.8);

\end{tikzpicture}
\caption{Acting with a topological interface we can relate a boundary condition with data $(p,v,k)$ in the theory at coupling $\tau$ to a new one with data $(p_f,v_f,k_f)$ in the theory at coupling $\tau'$.}
\label{fig: duality wall}
\end{figure}

Given the set of all possible free boundary conditions, we may now ask whether there exist interesting maps that relate them while preserving part of their structure. A natural operation on the space of boundary conditions arises by composing them with \emph{topological interfaces} that separate two bulk theories with couplings $\tau$ and $\tau'$, respectively. Since these interfaces are topological, they can be freely moved onto the boundary, thereby generating equivalences between boundary QFTs (see Fig.~\ref{fig: duality wall}).

As discussed in \cite{Diatlyk:2023fwf}, given a topological interface separating a QFT $\mathcal{T}$ from another $\mathcal{T}'$, one can show that there exists a symmetry action $\mathcal{S}$ such that
\begin{equation}
    \mathcal{T}' \;\cong\; \mathcal{T}/\mathcal{S}
\end{equation}
where $\mathcal{T}/\mathcal{S}$ denotes the theory obtained from $\mathcal{T}$ by gauging some generalized symmetry $\mathcal{S}$. The above isomorphism between $\mathcal{T}'$ and $\mathcal{T}/\mathcal{S}$ can be trivial, or it may implement a non-trivial duality of the theory. In particular, there can exist non-trivial interfaces between $\mathcal{T}$ and $\mathcal{T}' \cong \mathcal{T}$, which we refer to as \emph{duality interfaces}. Such interfaces act on the space of bulk observables as an automorphism (they merely implement the duality transformation), and they also provide a non-trivial action on the space of boundary conditions. In our case, the bulk Maxwell theory enjoys an $SL(2,\bZ)$ duality and a $U(1)\times U(1)$ 1-form symmetry, from which one can generate topological interfaces by gauging non-anomalous subgroups on half-space. In the following, we construct such topological interfaces and analyze their action on the set of free boundary conditions previously described.
\subsection{Duality interfaces}\label{sec: halfgauging int}

Let us start by constructing the topological $SL(2,\bZ)$ duality interfaces. As noticed in \cite{Paznokas:2025epc}, these interfaces can be constructed by gauging (with discrete torsion) trivial $\bZ_1 \times \bZ_1$ subgroups of the $U(1) \times U(1)$ 1-form symmetry of the theory. We will first start with the case $\theta=0$ for simplicity. For convenience, we use the following notation for the Maxwell action:
\begin{equation}\label{modified Maxwell theta=0}
S_{4d}=\int \frac{e^2}{4\pi}\Upsilon\wedge *\Upsilon+\frac{i}{2\pi}\Upsilon\wedge F\ .
\end{equation}
Integrating out $\Upsilon$ returns the Maxwell action, with $\theta=0$. 
The current of the electric symmetry in this notation is $*J_e=\frac{1}{2\pi}\Upsilon$. We consider the gauging of a $\bZ_1\times \bZ_1\subset U(1)_e\one\times U(1)_m\one$ symmetry of Maxwell as follows:\footnote{The introduction of $\Upsilon$ is a mathematical trick to isolate the $*$ and simplify later expressions. There is no physical meaning behind its introduction. In particular, we do not consider gauge transformations for this field.}
\begin{equation}\label{Maxwell Z1 gauging}
\begin{split}
    S_{4d}=\int \frac{e^2}{4\pi}\Upsilon\wedge *\Upsilon &+\frac{i}{2\pi}\Bigl(\Upsilon\wedge F+B\wedge(\gamma\Upsilon+\delta F+dA' )+C\wedge(\alpha \Upsilon+\beta F+d\Psi)\Bigr. \\
    &\Bigl.+\frac{\gamma\delta }{2}B \wedge B+\frac{\alpha \beta }{2}C\wedge C+\alpha \delta B\wedge C\Bigr)
\end{split}
\end{equation}
where $\alpha,\beta,\gamma,\delta\in\bZ$ and $\alpha \delta -\beta \gamma=1$ to ensure gauge invariance. Since $\gcd(\gamma,\delta )=1$ and $\gcd(\alpha ,\beta )=1$, $\gamma\Upsilon+\delta F$ and $\alpha \Upsilon+\beta F$ are two correctly normalized $U(1)$ currents. The condition $\alpha \delta -\beta \gamma=1$ imposes that these are currents of two independent symmetries. Here, the 2-form fields $B$ and $C$ play the role of gauge fields for the two symmetries, and we have introduced the $U(1)$ 1-form fields $A'$ and $\Psi$ as Lagrange multipliers which impose that $B$ and $C$ are $\bZ_1$ fields, i.e.~flat and with trivial holonomies.

The torsion terms, namely those quadratic in $B$ and $C$, are necessary for gauge invariance, since we have the gauge transformations
\begin{equation}
    B\to B+d\lambda_B\, , \qquad C\to C+d\lambda_C\, , \qquad A\to A+d\lambda_A-\gamma\lambda_B-\alpha \lambda_C\,,
\end{equation}
as well as those for $A' $ and $\Psi$ 
\begin{equation}
    A' \to A'  +d\lambda_{A' }\, , \qquad  \Psi\to \Psi+d\lambda_{\Psi}-\lambda_B  \,.
\end{equation}
The asymmetry of the above gauge transformations comes from the mixed 't Hooft anomaly between the two $U(1)$ symmetries of Maxwell. One of these two fields needs to have a non-trivial gauge transformation to cancel this anomaly. The field invariant under the gauge transformations of $B$ and $C$, here being $A' $, will become the new dynamical field of the gauged Maxwell, while the other will become an edge mode of the 3d interface between the gauged and ungauged Maxwell. Integrating out $B$ and $C$ gives
\begin{equation}
\begin{split}
    S_{4d} = \int &\frac{e^2}{4\pi}\Upsilon\wedge *\Upsilon +\frac{i}{2\pi}\left(-\frac{\gamma}{2\delta } \Upsilon\wedge \Upsilon-\frac{1}{\delta }\Upsilon \wedge dA' +\frac{\beta }{2\delta } dA'  \wedge dA' -dA'  \wedge dA'  \right. \\
    &\qquad\left. -\frac{\beta }{2\alpha }F \wedge F +\frac{\gamma}{2\alpha }d\Psi \wedge d\Psi-\frac{1}{\alpha }F \wedge  d\Psi\right)\,.
\end{split}
\end{equation}
Finally, integrating out $\Upsilon$ yields,
\begin{equation}\label{SL2Z Maxwell action}
\begin{split}
    S_{4d} = \int &\frac{1}{4\pi e^2}\frac{1}{\delta ^2+\left(\gamma\frac{1}{e^2}\right)^2}dA' \wedge *dA'  +\frac{i}{4\pi}\frac{\beta \delta  +\alpha \gamma\left(\frac{1}{e^2}\right)^2}{\delta ^2+\left(\gamma\frac{1}{e^2}\right)^2} dA'  \wedge dA'  \\
    &+\frac{i}{2\pi}d\left(-A'  d\Psi-\frac{\beta }{2\alpha }A\wedge dA +\frac{\gamma}{2\alpha }\Psi\wedge  d\Psi-\frac{1}{\alpha }A \wedge d\Psi\right)\,.
\end{split}
\end{equation}
The first two terms of \eqref{SL2Z Maxwell action} correspond to the Maxwell action where 
\begin{equation}
    \tau\to \frac{\alpha \tau+\beta }{\gamma\tau+\delta } = \tau'\,.
\end{equation}
The last term in brackets in \eqref{SL2Z Maxwell action} is a total derivative and produces a 3d action living at the interface between the two dual theories with $\tau$ and $\tau'$. At first glance, these boundary terms seem to be improperly quantized when $\alpha \neq 1$. However, it is possible to rewrite the action of this 3d interface in the following way:
\begin{equation}\label{3d 2edge modes}
    I_{3d}=\frac{i}{2\pi}\int_{\partial}-\alpha A'  \wedge d\Phi+ \beta A'  \wedge  dA+\frac{\beta \delta }{2}A\wedge dA
    +\frac{\alpha \gamma}{2}\Phi \wedge d\Phi-\alpha \delta A  \wedge d\Phi
\end{equation}
via the field redefinition,
\begin{equation}\label{renormalized edge mode}
\Phi:=\frac{1}{\alpha }\Psi+\frac{\beta }{\alpha }A\,.
\end{equation}
The gauge transformations of $\Phi$ are:
\begin{equation}\label{renormalized edge mode gauge}
    \Phi\to \Phi-\delta \lambda_B-\beta \lambda_C\ .
\end{equation}
The coefficients in front of $\lambda_B$ and $\lambda_C$ are both integers despite the division by $\alpha $ in the field redefinition \eqref{renormalized edge mode}.\footnote{Note that for this field redefinition and the integration of $B$ and $C$ above, we supposed that $\alpha\neq 0$. However, the interface action \eqref{3d 2edge modes} still holds in the case $\alpha=0$. In this case, the field $\Psi$ disappears from the final expression.} This guarantees that $\Phi$ has well-defined gauge transformations, thus validating such a field redefinition.

The case $\theta\neq 0$ can be treated in the similar way. Introducing a term $\frac{\theta}{4\pi}FF$ to \eqref{modified Maxwell theta=0}, leads to the action\footnote{One could alternatively consider the following bulk action \begin{equation}\label{modified Maxwell 2}
\int \frac{1}{4\pi e^2 |\tau|^2}\Upsilon\wedge *\Upsilon-\frac{i\theta}{8\pi^2 |\tau|^2}\Upsilon\wedge \Upsilon+\frac{i}{2\pi} \Upsilon \wedge F\ .
\end{equation}
With this choice, the current of the electric field remains $*J_e=\frac{1}{2\pi}\Upsilon$. Starting with \eqref{modified Maxwell 2} or \eqref{modified Maxwell} does not affect the final result as both are two equivalent ways to write the same action.
}
\begin{equation}\label{modified Maxwell}
S_{4d}=\int \frac{e^2}{4\pi}\Upsilon\wedge *\Upsilon+\frac{i}{2\pi}\left(\Upsilon\wedge F+\frac{\theta}{4\pi}FF\right)\,.
\end{equation}
This term modifies the current of the electric symmetry which is now:
\begin{equation}
    *J_e=\frac{1}{2\pi}\left( \Upsilon+\frac{\theta}{2\pi}F\right)\ .
\end{equation}
This modified current affects the gauging of the $\bZ_1\times \bZ_1$, and the action \eqref{Maxwell Z1 gauging} needs to be adapted as follows:
\begin{equation}\label{Maxwell Z1 gauging theta}
\begin{split}
\int &\; \frac{e^{2}}{4\pi}\,\Upsilon \wedge *\Upsilon 
+ \frac{i}{2\pi} \Bigg(
   \Upsilon \wedge F 
   + \frac{\theta}{4\pi} F \wedge F 
   + B \wedge \big( \gamma \Upsilon + (\delta + \tfrac{\theta}{2\pi}\gamma) F + dA' \big) \\
&\quad 
   + C \wedge \big( \alpha \Upsilon + (\beta + \tfrac{\theta}{2\pi}\alpha) F + d\Psi \big)
   + \tfrac{\gamma(\delta + \tfrac{\theta}{2\pi}\gamma)}{2} \, B \wedge B
   + \tfrac{\alpha(\beta + \tfrac{\theta}{2\pi}\alpha)}{2} \, C \wedge C \\
&\quad 
   + \alpha (\delta + \tfrac{\theta}{2\pi}\gamma) \, B \wedge C
\Bigg) .
\end{split}
\end{equation}
Additionally, gauge invariance now requires the following modification of the parameters in the torsion terms:
\begin{equation}
    \beta \to \beta +\frac{\theta}{2\pi}\alpha \, ,\qquad \delta \to \delta +\frac{\theta}{2\pi}\gamma\ .
\end{equation}
The gauge transformations of the fields remain the same as in the case where $\theta=0$. Since the action with $\theta\neq 0$ differs from the action with $\theta = 0$, simply by a total derivative and the above parameter redefinition, we can directly conclude from the case $\theta=0$ that the transformation of $\tau$ will be:
\begin{equation}
   \tau= \frac{i}{e^2}+\frac{\theta}{2\pi}\to \frac{\alpha \frac{i}{e^2}+(\beta+\frac{\theta}{2\pi}\alpha)}{\gamma \frac{i}{e^2}+(\delta+\frac{\theta}{2\pi}\gamma)}=\frac{\alpha \left(\frac{i}{e^2}+\frac{\theta}{2\pi}\right)+\beta}{\gamma \left(\frac{i}{e^2}+\frac{\theta}{2\pi}\right)+\delta}=\frac{\alpha \tau+\beta}{\gamma \tau+\delta}\ .
\end{equation}
The total derivative terms that will define the 3d interface are
\begin{equation}
    I_{3d}=\frac{i}{2\pi}\int -A'  d\Psi-\frac{(\beta+\frac{\theta}{2\pi}\alpha) }{2\alpha }A\wedge dA +\frac{\gamma}{2\alpha }\Psi\wedge  d\Psi-\frac{1}{\alpha }A \wedge d\Psi+\frac{\theta}{4\pi}AdA\,.
\end{equation}
We observe that the modification of $\beta$ cancels the $\frac{\theta}{4\pi}FF$ term that we added in the bulk and we are left with the same 3d interface as in the case $\theta=0$. Note that we still need to use the redefinition $\Phi=\frac{1}{\alpha}(\Psi+\beta A)$ with the unmodified $\beta$ in this case, otherwise $\Phi$ would have not well-defined gauge transformations. The 3d interface between the initial and gauged theories therefore remains \eqref{3d 2edge modes}. 

We can now consider fusing the 3d interface \eqref{3d 2edge modes} with a general 3d free boundary theory of the ungauged Maxwell. After such fusion, the original Maxwell field $A$ now only lives on the 3d boundary and is therefore interpreted as an additional edge mode. The new parameters and fields of the fused boundary are
\begin{equation}
    k_f=\begin{pmatrix}
       k & v & 0_{n+1}\\
       v^T & \beta \delta  & -\alpha \delta \\
       0_{n+1}^T & -\alpha \delta  & \alpha \gamma
    \end{pmatrix} \, , \quad v_f=\begin{pmatrix}
        0_{n+1}\\
        \beta \\
        -\alpha 
    \end{pmatrix} \, , \quad \Phi_{f,i}=\begin{pmatrix}
        \Phi_i^{3d}\\
        A\\
        \Phi
    \end{pmatrix}\ ,
\end{equation}
where $0_{n+1}$ is a column vector of $(n+1)$ zeros and $\Phi_i^{3d}$ are the $(n+1)$ edge modes of the initial boundary. The new boundary conditions are characterized by 
\begin{equation}\label{eq: SL2z on P,Q}
    \frac{Q_f}{P_f}:=v_f^Tk_f^{-1}v_f=\frac{\alpha Q+\beta P}{\gamma Q  +\delta P},
\end{equation}
where we used $v^Tk^{-1}v=\frac{Q}{P}$ and the following inverse for $k_f$:
\begin{equation} \label{kf inverse}
    k^{-1}_f = \begin{pmatrix} k^{-1}-\frac{\gamma P}{\delta P+\gamma Q}k^{-1}vv^Tk^{-1} & k^{-1} v \frac{\gamma P}{\delta P +\gamma Q} & k^{-1}v \frac{\delta P }{\delta P+\gamma Q}\\
    v^T k^{-1}  \frac{\gamma P}{\delta P +\gamma Q}& \frac{-\gamma P }{\delta P + \gamma Q} & \frac{-\delta P }{\delta P +\gamma Q} \\ 
    v^T k^{-1}\frac{\delta P }{\delta P+\gamma Q} & \frac{-\delta P}{\delta P +\gamma Q} & \frac{-\beta \delta P + Q}{\alpha (\delta P +\gamma Q)}
    \end{pmatrix}\,.
\end{equation}
We conclude that the $\bZ_1\times \bZ_1$ gauging not only implements an $SL(2,\bZ)$ transformation of the 4d bulk action but also implements an $SL(2,\bZ)$ transformation of the boundary conditions. In particular, this trivial gauging does not alter the set of operators that can terminate on the 3d boundary but simply corresponds to a relabeling of these operators, as we now show.

In order to determine the global boundary condition, we use that the combination $k^{-1}v$ after fusing with the defect becomes
\begin{equation}
    k_f^{-1}v_f=\frac{1}{\gamma Q  +\delta P}\begin{pmatrix}
        - P k^{-1}v\\
        P\\
        -Q
    \end{pmatrix}\,.
\end{equation}
Since $\gcd(P,Q)=1$, we eventually get that $R=rP$ now becomes $R_f=r(\gamma Q  +\delta P)$, which implies $r_f=r$, and $g_f=\gcd(g,rP,rQ)=\gcd(g,r)$. In particular, we have 
\begin{equation}
    \tilde r=\frac{r}{\gcd(g,r)}=\frac{r_f}{\gcd(g_f,r_f)}=\tilde r_f\,,
\end{equation}
and new operators trivialized by the boundary conditions are
\begin{equation}
    (W_{\alpha Q+\beta P }H_{\gamma Q  +\delta P}(\gamma))^{\tilde r m}\quad \forall  m\in \bZ\,.
\end{equation}
The set of lines trivialized by the duality-transformed boundary condition coincides with the original one. Therefore, we conclude that the duality interface does not change the global structure of the boundary conditions. 

By choosing the reference boundary to be the one with no boundary edge modes, after fusing it with the duality interface we get a boundary TQFT with only two edge modes, such that $v_f^Tk^{-1}_fv_f=\frac{\beta }{\delta }$ and $\tilde r=1$. This implies that the class of boundary actions given by \eqref{3d 2edge modes} spans all the possible boundary conditions with $\tilde r=1$, related to each other through an $SL(2,\bZ)$ transformation. Incidentally, we deduce that the boundary with no edge modes is associated to Neumann with $\tilde r=1$.\footnote{It is actually possible to write more economical boundary actions with a single edge mode, for instance 
\begin{equation}
    \frac{i}{2\pi}\int \tilde{r} Q A\wedge d\Phi+\frac{\tilde{r}^2QP}{2}\Phi\wedge d\Phi\ ,
\end{equation}
even allowing for a general $\tilde r$ (though this class does not encompass Neumann with $\tilde r\neq 1$). However, we find that the theory with two edge modes is better suited to our discussion.}

\subsubsection{S and T Transformations}
We now specialize the above general discussion to the case of $S$ and $T$ transformations. We hope to illustrate the method used, as well as highlight some important takeaways which we shall refer back to later when discussing the SymTFT setup.

Let us begin with the $T$ transformation,
\begin{equation}
    \begin{pmatrix}
        \alpha & \beta \\ \gamma & \delta \end{pmatrix} = \begin{pmatrix}
            1 & 1 \\ 0 & 1
        \end{pmatrix}
    \,,
\end{equation}
which is implemented by the following gauging,
\begin{equation}
    S_{4d}^T=\int \frac{e^2}{4\pi}\Upsilon\wedge *\Upsilon +\frac{i}{2\pi}\Bigl(\Upsilon\wedge F+B\wedge(F+dA' )+C\wedge( \Upsilon+ F+d\Psi)- \frac{1}{2}C\wedge C+ B\wedge C\Bigr)\ .
\end{equation}
When performed on half-spacetime, this yields the 3d interface,
\begin{equation}
    I_{3d}^T = \frac{i}{2\pi } \int_\partial -A'  \wedge d \Phi +A'   \wedge dA +\frac{1}{2} A\wedge dA -A \wedge d \Phi\,.
\end{equation}
Integrating out $\Phi$ imposes the constraint $A=-A' $, and, we are left with:
\begin{equation}
    I_{3d}^T = -\frac{i}{4\pi}\int_\partial A \wedge dA\,,
\end{equation}
which is consistent with the $T$ interface in \cite{Kapustin:2009av}.
After fusing the interface with the existing 3d boundary, the resulting 3d boundary is described by the parameters
\begin{equation}
    Q_f= Q+P \quad , \quad P_f =P\,.
\end{equation}

Similarly, for the $S$ transformation, which is given by the $SL(2,\bZ)$ matrix:
\begin{equation}
    \begin{pmatrix}
        \alpha & \beta \\ \gamma & \delta \end{pmatrix} = \begin{pmatrix}
            0& 1 \\ -1 & 0
        \end{pmatrix}
    \,.
\end{equation}
We find the $S$ transformation is given by the following gauging:
\begin{equation}
    S_{4d}^S = \int \frac{e^2}{4\pi}\Upsilon \wedge * \Upsilon +\frac{i}{2\pi} \left(\Upsilon \wedge F + B \wedge (-\Upsilon+dA'  ) +C \wedge (F+d\Psi )\right) \,.
\end{equation}
The resulting 3d interface in this case is,
\begin{equation}
    I_{3d}^S = \frac{i}{2\pi} \int_\partial A'  \wedge  dA \,,
\end{equation}
again in line with \cite{Kapustin:2009av}.
Fusing this interface with the 3d boundary produces a new 3d boundary described via:
\begin{equation}
    Q_f=P \quad ,\quad P_f =-Q\,.
\end{equation}

\subsection{Rescaling interfaces}

A different operation that we can perform in the 4d bulk is to gauge a $\bZ_\cN\times \bZ_\cM$ subgroup of $U(1)_e\one\times U(1)_m\one$. For simplicity, we take $\bZ_\cN\subset U(1)_e\one$ and $\bZ_\cM\subset U(1)_m\one$, with of course $\gcd(\cN,\cM)=1$ to ensure that there is no mixed anomaly. We expect the result of such an operation to be a rescaling of the coupling $e^2\to \frac{\cN^2}{\cM^2}e^2$. Below, we implement such $\bZ_\cN\times \bZ_\cM$ gauging in half of spacetime, in order to determine the resulting topological interface. Eventually, we want to see how this interface fuses with the edge mode theory that implements the boundary conditions. We will see that it modifies their global properties.

It turns out that a simple way to do the $\bZ_\cN\times \bZ_\cM$ gauging is the following:
\begin{equation}\label{Maxwell ZNZM gauging}
    S_{4d}=\int \frac{e^2}{4\pi}\Upsilon\wedge *\Upsilon +\frac{i}{2\pi}\Bigl(\Upsilon\wedge F+B\wedge(\cN F+\cM dA' )+C\wedge( \Upsilon+\cN d\Psi)+\cN B\wedge C\Bigr)\ ,
\end{equation}
where $F$ and $\Upsilon$ are the magnetic and electric currents respectively, and we have introduced the $U(1)$ fields $A'$ and $\Psi$ as Lagrange multipliers imposing that $B\,,C$ are $\bZ_\mathcal{M}$ and $\bZ_\mathcal{N}$ fields. Note that we have somewhat unconventionally written the coupling of $B$ to the magnetic current with an $\cN$ in front to ensure gauge invariance, but the fact that $\gcd(\cN,\cM)=1$ still ensures that we are correctly gauging a $\bZ_\cM$ subgroup.\footnote{By interpreting a discrete gauging as the sum over symmetry operators wrapping non-trivial cycles, one finds that summing over insertions of $\exp \left(\frac{i k \mathcal{N}}{\mathcal{M}}\int F\right)$, with $k\in \bZ_{\mathcal{M}}$, is equivalent to summing over insertions of $\exp \left(\frac{i k}{\mathcal{M}}\int F\right)$ operators if $\gcd(\mathcal{N},\mathcal{M})=1$.}
The fields have the gauge transformations
\begin{equation}
    A \to A+d\lambda_A -\lambda_C \quad , \quad A'  \to A' +d\lambda_{A' } \quad ,\quad \Psi \to \Psi+d\lambda_{\Psi} -\lambda_B\ ,
\end{equation}
such that the above action is gauge invariant. 

Integrating out $B$ and $C$ gives 
\begin{equation}
    S_{4d}=\int \frac{e^2}{4\pi}\Upsilon\wedge *\Upsilon -\frac{i}{2\pi}\Bigl(\frac \cM\cN \Upsilon\wedge dA' +\cN F\wedge d \Psi+ \cM dA'  \wedge d\Psi \Bigr)\ ,
\end{equation}
so that after further integrating out $\Upsilon$ we get
\begin{equation}
    \int \frac{\cM^2}{4\pi \cN^2 e^2}  dA'  \wedge * dA'  -\frac{i}{2\pi} (\cN F\wedge d\Psi +\cM dA'  \wedge d\Psi ) \,.
\end{equation}
Notice that $dA' $ is already a good field strength for the new Maxwell action since it is gauge invariant, and it has exactly the Maxwell action with rescaled coupling that we expected. 

The last two terms eventually define the 3d interface between the Maxwell theories defined with $\tau$ and $\tau'$
\begin{equation}\label{eq: rascaling interface}
    I_{3d}= -\frac{i}{2\pi} \int \cN A \wedge d\Psi +\cM A' \wedge  d\Psi \,.
\end{equation}
We can then fuse this interface with the existing 3d boundary, to yield
\begin{equation}
    k_f =\begin{pmatrix}
        k& v & 0_{n+1} \\ v^T & 0 & -\cN \\ 0^T_{n+1} & -\cN & 0
    \end{pmatrix} \quad , \quad v_f =\begin{pmatrix}
        0_{n+1} \\ 0 \\-\cM
    \end{pmatrix} \quad , \quad \Phi_{f,i} = \begin{pmatrix}
        \Phi^{3d}_i \\ A \\ \Psi
    \end{pmatrix}
\end{equation}
such that 
\begin{equation}
    k^{-1}_f v_f = \begin{pmatrix}
        -k^{-1}v \frac{\cM}{\cN} \\ \frac{\cM}{\cN} \\ -\frac{Q\cM}{P\cN^2}
    \end{pmatrix}
\end{equation}
and 
\begin{equation} \label{rescaled vkv}
    v^T_f k^{-1}_f v_f =\frac{Q}{P} \frac{\cM^2}{\cN^2}\,.
\end{equation}
This allows us to deduce that 
\begin{equation}
    Q_f=\frac{Q\cM^2}{\gcd(Q\cM^2,P\cN^2) }\quad,\quad P_f =\frac{P\cN^2}{\gcd(Q\cM^2,P\cN^2)}\,.
\end{equation}
An integer $R_f$ such that $R_f k_f^{-1}v_f$ are all integers is straightforwardly given by $R_f = rP\cN^2$. It can be rewritten as $R_f=r_f P_f$, such that $r_f = r \gcd (Q\cM^2,P\cN^2)$. We then have $g_f = \gcd(g\cM\cN,rP\cM\cN,rQ\cM)$. 
The quantity of interest is then
\begin{equation}
    \gcd(r_f,g_f)=\gcd(g\cM\cN,rP\cM\cN,rQ\cM,rQ\cM^2,rP\cN^2)\ .
\end{equation}
Exploiting the fact that $\gcd(rQ\cM,rQ\cM^2)=rQ\cM$ and $\gcd(rP\cM\cN,rP\cN^2)=rP\cN$ (due to $\gcd(\cM,\cN)=1$), we simplify this to
\begin{equation}
    \gcd(r_f,g_f)=\gcd(g\cM\cN,rP\cN,rQ\cM)=\gcd(g\cM\cN,r\gcd(P\cN,Q\cM))\ .
\end{equation}
By using the fact that $\gcd(P,Q) = 1 = \gcd(\cM,\cN)$, it follows that $\gcd(Q\cM^2,P\cN^2)=\gcd(Q,\cN^2)\gcd(P,\cM^2) $ and  $\gcd(Q\cM,P\cN)=\gcd(Q,\cN)\gcd(P,\cM) $. Finally, the relevant ratio can be written as
\begin{equation}\label{tilde r rescaling}
    \tilde r_f= \tilde r \frac{\gcd( P, \cM^2)\gcd( Q, \cN^2)}{\gcd( \cM \cN,\tilde r\gcd(P,\cM)\gcd(Q,\cN))}\ . 
\end{equation}
The initial global boundary conditions trivialize the following operators
\begin{equation}
    (W_Q H_P)^{\tilde r m} =1 \quad m \in \bZ\ .
\end{equation}
From the above, we observe that after the $\bZ_\cN^e \times \bZ_\cM^m$ gauging, the new global boundary conditions become
\begin{equation}
    (W_{Q_f} H_{P_f})^{\tilde r_f m} =1 \quad m \in \bZ\ .
\end{equation}
Since the above expression is still a bit implicit, we are going to show a few simple examples where it will become evident that, after the gauging, the quantity $\tilde r_f$ can actually increase or decrease.

First of all, one can easily check that if $\cN=\cM=1$ (trivial gauging), then all quantities remain unchanged, in particular $\tilde r_f=\tilde r$.

For a situation where $\tilde r_f$ is larger than $\tilde r$, take the latter to be minimal, $\tilde r=1$, then 
\begin{equation}
    \tilde r_f=\gcd\left(\frac{P}{\gcd(P,\cM)}, \cM\right)\gcd\left(\frac{Q}{\gcd(Q,\cN)}, \cN\right)\, .
\end{equation} 
This quantity is maximal if we consider $P=\cM^2$, $Q=\cN^2$, so that $\tilde r_f=\cM\cN$.
In this case, the trivial lines are rearranged as
\begin{equation}
    (W_{\cN^2}H_{\cM^2})^\ell=1 \quad \to \quad (W_1H_1)^{\cM\cN \ell }=1 \qquad \ell \in \bZ\ .
\end{equation}

For the opposite case in which $\tilde r_f$ is smaller than $\tilde r$, let us take $\gcd(P,\cM)=\gcd(Q,\cN)=1$, and $\tilde r=\cM\cN$. Then $\tilde r_f=1$, and the lines\footnote{Notice that for any $P,Q,\tilde r$ we can always find a decomposition $\tilde r = \cM\cN$ such that this condition is satisfied. Therefore, for any non-trivial $\tilde r$ we can always select a topological interface which trivializes it. This is in accordance with the fact that the $\bZ_{\tilde r}$ 1-form symmetry is not anomalous.} rearrange as
\begin{equation}
    (W_Q H_P)^{\cM\cN\ell}=1 \quad \to \quad (W_{Q\cM^2}H_{P\cN^2})^\ell =1 \quad \ell \in \bZ\ .
\end{equation}
Note that for $Q=P=1$ this gauging has the opposite effect as the previous one (interchanging the roles of $\cN$ and $\cM$). Other examples can be treated in a similar manner.

\subsubsection{Condensation defect}
\label{sec: condensation defect}
We can actually consider performing a $\bZ_\cN^e \times \bZ_\cM^m$ gauging, followed by a $\bZ_\cM^e \times \bZ_\cN^m$ gauging. The two gaugings result in a trivial rescaling of the coupling. Indeed, after both gaugings, we recover the initial bulk theory, however, generically with a different spectrum of endable operators. The 3d interface for the second gauging is given by
\begin{equation}
    I_{3d}' = -\frac{i}{2\pi} \int \cM A' \wedge d\Psi' +\cN \hat{A}\wedge d\Psi'\,,
\end{equation}
where $\hat{A}$ is the gauge field on the left of the two interfaces, while $A'$ now lives in the slab between them.

Fusing both of these gauging interfaces with the existing 3d boundary, the 3d edge mode theory is now described by the parameters (which are denoted with a hat)
\begin{equation}
    \hat{k} = \begin{pmatrix}
        k_f & v_f & 0 \\
        v_f^T & 0 & -\cM \\
        0^T & -\cM & 0
    \end{pmatrix} \quad , \quad \hat{v} = \begin{pmatrix}
        0 \\ 0 \\ -\cN
    \end{pmatrix}\, \quad , \quad \hat \Phi_i = \begin{pmatrix}
        \Phi_i \\ A' \\ \Psi'
    \end{pmatrix}= \begin{pmatrix}
        \Phi_i^{3d} \\ A \\ \Psi \\ A' \\ \Psi'
    \end{pmatrix}\ .
\end{equation}
One then finds that
\begin{equation}
    \hat{k}^{-1} \hat{v} = \begin{pmatrix}
        -k^{-1}_f v_f \frac{\cN}{\cM} \\ \frac{\cN}{\cM} \\ -\frac{Q}{P\cN} \\ 
    \end{pmatrix} = \begin{pmatrix}
        k^{-1} v \\ -1 \\ \frac{Q}{P\cN} \\ \frac{\cN}{\cM} \\ -\frac{Q}{P\cN}
    \end{pmatrix}\, ,
\end{equation}
such that 
\begin{equation}
    \frac{\hat{Q}}{\hat{P}}=\hat{v}^T \hat{k}^{_1} \hat{v} \equiv \frac{Q}{P}\,.
\end{equation}
Thus, we find that the local boundary condition is unchanged after composing the $\bZ_\cN^e \times \bZ_\cM^m$ gauging with $\bZ_\cM^e \times \bZ_\cN^m$ gauging as we expect. Now, turning to the global form of the boundary condition, the above implies that $\hat{R} = \hat{r}P = rP\cN\cM$ and $\hat{g} = \gcd(g\cN\cM,rP\cN,rQ\cM)$. We are interested in the gcd:
\begin{equation}
    \begin{split}
        \gcd(\hat{g},\hat{r}) &= \gcd(g\cN\cM,rP\cN,rQ\cM,r\cN\cM) \\ &= \gcd(g,r) \gcd(\cN\cM , \tilde{r}\gcd(P\cN,Q\cM)) 
    \end{split}
\end{equation}
where $\tilde r$ is defined as before. The endable lines in this case are
\begin{equation}\label{boundary condition non-invertible defect}
    (W_Q H_P)^{\tilde r \frac{\cN\cM}{\gcd(\cN\cM,\tilde{r}\gcd(P\cN,Q\cM))} \ell } \quad \ell \in \bZ\,.
\end{equation}
We can see that generically less lines are trivialized on the boundary after this double gauging. 

For instance, take $\tilde r=1$ to start with. If $P$ and $Q$ are coprime also with $\cM$ and $\cN$, respectively, then the effect of the double gauging is to go from $W_QH_P=1$ to $(W_QH_P)^{\cM\cN}=1$, namely we end up with much fewer trivial lines. If on the other hand $P=\cM$, $Q=\cN$ then the set of trivial lines is unchanged.

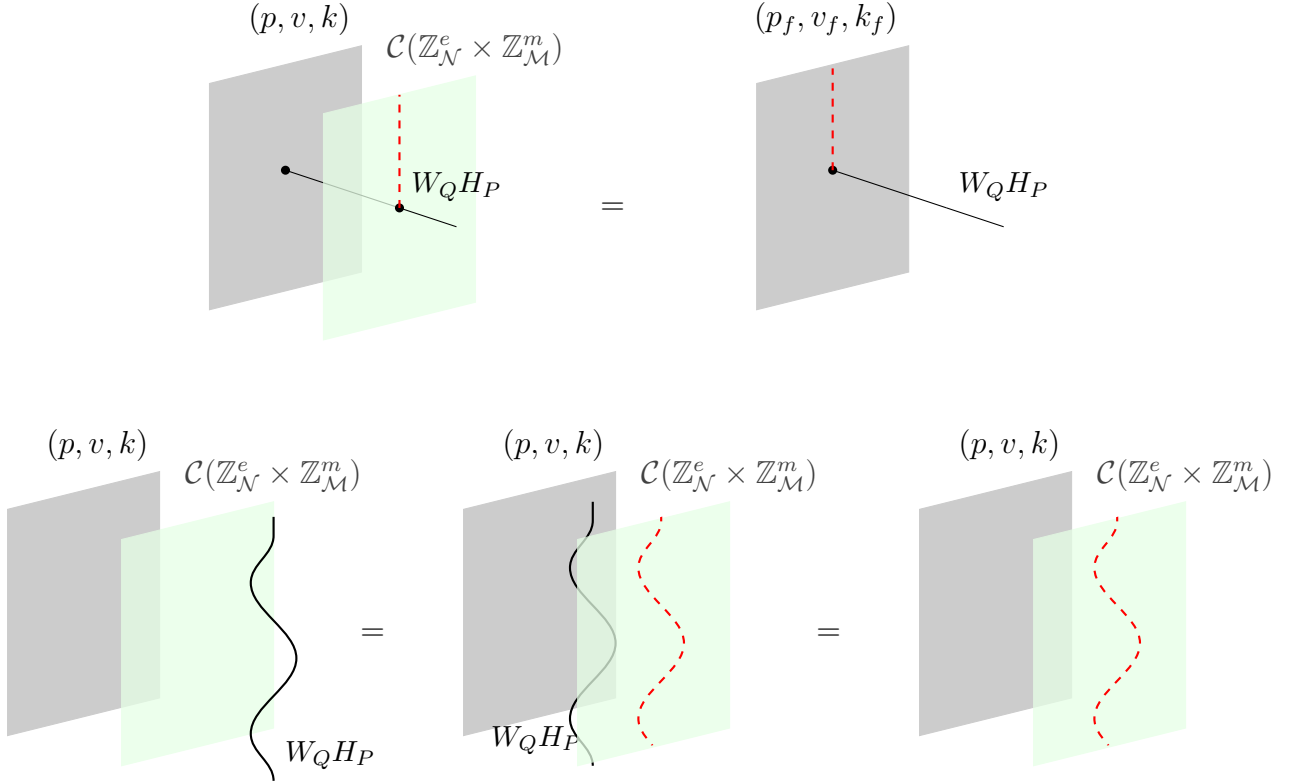
\begin{figure}[t]
\centering

\begin{tikzpicture}

\coordinate (in) at (2.5,1.25); 
\coordinate (inu) at (2.5,1.5); 
\coordinate (ind) at (2.5,1); 
\coordinate (out) at (-0.5,2.25); 
\coordinate (def) at (1,1.75);

\begin{scope}[shift={(-1.5,0.4)}]
    \filldraw[color=white!80!black] 
        (0,0) -- (2,0.5) -- (2,3.5) node[above left, color=black] {$(p,v,k)$} -- (0,3) -- cycle;
\end{scope}

\draw[color=black] (def) to (out) node[left] {} node[circle, fill, inner sep=1.2pt] {};

\filldraw[color=white!90!green, opacity=0.75] 
    (0,0) -- (2,0.5) -- (2,3.5) node[above, color=black] {$\mathcal{C}(\mathbb{Z}_\cN^e\times \mathbb{Z}_\cM^m)$} -- (0,3) -- cycle;

\draw[color=black] ($0.5*(in)+0.5*(def)$) node[above=0.2cm] {\small $W_QH_P$} to (def) node[circle, fill, inner sep=1.2pt] {};

\draw[color=red, thick, dashed] (def) to (1,3.25);

\node[right] at (3.5,1.75) {$=$};

\begin{scope}[shift={(7.2,0)}]
    \coordinate (in) at (2.5,1.25); 
    \coordinate (inu) at (2.5,1.5); 
    \coordinate (ind) at (2.5,1); 
    \coordinate (out) at (-0.5,2.25); 
    \coordinate (def) at (1,1.75);

    \begin{scope}[shift={(-1.5,0.4)}]
        \filldraw[color=white!80!black] 
            (0,0) -- (2,0.5) -- (2,3.5) node[above left, color=black] {$(p_f,v_f,k_f)$} -- (0,3) -- cycle;
    \end{scope}

    \draw[color=black] (def) to (out) node[left] {} node[circle, fill, inner sep=1.2pt] {};
    \draw[color=black] ($0.5*(in)+0.5*(def)$) node[above=0.2cm] {\small $W_QH_P$} to (def);
    \draw[color=red, thick, dashed] (out) to (-0.5,3.6);
\end{scope}

\end{tikzpicture}

\vspace{1cm} 

\begin{tikzpicture}

\coordinate (in) at (2.5,1.25); 
\coordinate (inu) at (2.5,1.5); 
\coordinate (ind) at (2.5,1); 
\coordinate (out) at (-0.5,2.25); 
\coordinate (def) at (1,1.75);

\begin{scope}[shift={(-1.5,0.4)}]
    \filldraw[color=white!80!black] 
        (0,0) -- (2,0.5) -- (2,3.5) node[above left, color=black] {$(p,v,k)$} -- (0,3) -- cycle;
\end{scope}

\filldraw[color=white!90!green, opacity=0.75] 
    (0,0) -- (2,0.5) -- (2,3.5) node[above, color=black] {$\mathcal{C}(\mathbb{Z}_\cN^e\times \mathbb{Z}_\cM^m)$} -- (0,3) -- cycle;

\begin{scope}[shift={(0.8,-0.2)}]
    \draw[decorate, decoration={snake, amplitude=3mm, segment length=20mm}, thick, black] 
        (1.2,0.) node[above right, color=black] {\small $W_QH_P$} -- (1.2,3.5);
\end{scope}

\node[right] at (3.,1.75) {$=$};

\begin{scope}[shift={(6.,0)}]
    \coordinate (in) at (2.5,1.25); 
    \coordinate (inu) at (2.5,1.5); 
    \coordinate (ind) at (2.5,1); 
    \coordinate (out) at (-0.5,2.25); 
    \coordinate (def) at (1,1.75);

    \begin{scope}[shift={(-1.5,0.4)}]
        \filldraw[color=white!80!black] 
            (0,0) -- (2,0.5) -- (2,3.5) node[above left, color=black] {$(p,v,k)$} -- (0,3) -- cycle;
    \end{scope}

    \begin{scope}[shift={(-1,0)}]
        \draw[decorate, decoration={snake, amplitude=3mm, segment length=20mm}, thick, black] 
            (1.2,0.) node[above left, color=black] {\small $W_QH_P$} -- (1.2,3.5);
    \end{scope}

    \filldraw[color=white!90!green, opacity=0.75] 
        (0,0) -- (2,0.5) -- (2,3.5) node[above, color=black] {$\mathcal{C}(\mathbb{Z}_\cN^e\times \mathbb{Z}_\cM^m)$} -- (0,3) -- cycle;

    \begin{scope}[shift={(-0.1,0)}]
        \clip (0,0) -- (2,0.5) -- (2,3.5) -- (0,3) -- cycle;
        \draw[decorate, decoration={snake, amplitude=3mm, segment length=20mm}, thick, red, dashed] 
            (1.2,0.) -- (1.2,3.5);
    \end{scope}

    \node[right] at (3.,1.75) {$=$};
\end{scope}

\begin{scope}[shift={(12.,0)}]
    \coordinate (in) at (2.5,1.25); 
    \coordinate (inu) at (2.5,1.5); 
    \coordinate (ind) at (2.5,1); 
    \coordinate (out) at (-0.5,2.25); 
    \coordinate (def) at (1,1.75);

    \begin{scope}[shift={(-1.5,0.4)}]
        \filldraw[color=white!80!black] 
            (0,0) -- (2,0.5) -- (2,3.5) node[above left, color=black] {$(p,v,k)$} -- (0,3) -- cycle;
    \end{scope}

    \filldraw[color=white!90!green, opacity=0.75] 
        (0,0) -- (2,0.5) -- (2,3.5) node[above, color=black] {$\mathcal{C}(\mathbb{Z}_\cN^e\times \mathbb{Z}_\cM^m)$} -- (0,3) -- cycle;

    \begin{scope}[shift={(-0.1,0)}]
        \clip (0,0) -- (2,0.5) -- (2,3.5) -- (0,3) -- cycle;
        \draw[decorate, decoration={snake, amplitude=3mm, segment length=20mm}, thick, red, dashed] 
            (1.2,0.) -- (1.2,3.5);
    \end{scope}
\end{scope}

\end{tikzpicture}

\caption{Top: How bulk lines ending on a particular boundary condition (gray) can become non-endable after the action of a condensation defect (green). Bottom: Similarly, bulk lines trivialized by the same boundary conditions become non-trivial, but topological (dashed red) after acting with the condensation defect.}
\label{fig: cond. defect}
\end{figure}

The intuition for this behavior is as follows. Note, first of all, that the fusion of the two interfaces defines a condensation defect for the $\bZ_\cN^e \times \bZ_\cM^m$ symmetry of the bulk theory \cite{Roumpedakis:2022aik}. Such a defect is non-invertible, and its effect is to trivialize the symmetry operators. As a result,  the condensation defect acts non-trivially on operators charged under the gauged symmetry. When the condensation defect crosses such line operator, a non-trivial topological line operator is inserted on the defect (a similar effect was observed in \cite{Arbalestrier:2024oqg,Arbalestrier:2025poq}). It is this line operator that is no longer trivialized on the boundary, unless its charge matches the order of the gauged symmetry (see Figure \ref{fig: cond. defect}). For example, if we consider the condensation defect
\begin{equation}\label{footnotewilson}
    i\int \Phi'\wedge \left(\frac{1}{2\pi}\cN d\Psi+*J_e\right)
\end{equation}
and move this defect across a Wilson line, of charge $q$, we will generate an operator $\exp\left(i q\int \Phi' \right)$. If $q\notin  \cN\bZ$, the introduction of this operator is required to allow $\int *J_e=q$ when integrated on a closed surface within the condensation defect that links with the Wilson line.
If we insert a Wilson line of charge $q$ on the boundary, it may be trivialized by the initial boundary condition, but the operator $\exp\left(i q\int \Phi' \right)$ living on the condensation defect is not necessarily trivialized. Since $\exp\left(i\cN q\int \Phi' \right)=1$, the operator $\exp\left(i q\int \Phi' \right)$ is necessarily a $\bZ_\cN$ phase.

More specifically, the 3d interface obtained after the two gaugings is $I_{3d}+I_{3d}'$
\begin{align} \label{double gauging}
    -\frac{i}{2\pi}\int &\Big[\, 
        \cN \hat{A}\wedge d\Psi' 
        + \cM A'\wedge d\Psi' 
        + \cM A'\wedge d\Psi 
        + \cN A\wedge d\Psi 
    \Big] \nonumber \\[4pt]
    &\;\;\longrightarrow\;
    -\frac{i}{2\pi}\int \Big[\, 
        \cN\, d\Psi'\wedge(\hat{A}-A) 
        + \Psi\wedge\!\big(\cM\, dA' + \cN\, dA\big)
    \Big] .
\end{align}
where we have redefined $\Psi\to \Psi-\Psi'$. The last two terms correspond to a condensation defect for a $\bZ_\cM$ subgroup of the magnetic symmetry (given that $\gcd(\cN,\cM)=1$). Its action on a 't Hooft line is
\begin{equation}
    H_P\to H_P'=H_P \exp\left(i\frac{P\cN}{\cM}\int \cM\Psi\right)
\end{equation}
with $\int \cM\Psi \in 2\pi \bZ$.

The first term in \eqref{double gauging} is equivalent to the condensation defect for a $\bZ_\cN$ subgroup of the electric symmetry \eqref{footnotewilson}. Indeed, consider the interface
\begin{equation}
    \frac{i}{2\pi}\int \cN d\Psi'\wedge (A-\hat{A})\ .
\end{equation}
Varying with respect to $A$ and $\hat{A}$ respectively gives the following boundary constraints:
\begin{equation}
    *J_e=-\frac{1}{2\pi}\cN d\Psi'\, ,\quad *J'_e=-\frac{1}{2\pi}\cN d\Psi'\ .
\end{equation}
Summing over holonomies of $\Psi'$ implies
\begin{equation}
    \hat{A}=A-\Phi'\,,
\end{equation}
where $\Phi'$ is some closed $U(1)$ field satisfying $\int \Phi'\in \frac{2\pi}{\cN}\bZ$. These two conditions, on the quantization of $*J_e$ and $\Phi'$, can be equivalently obtained 
from \eqref{footnotewilson}.
The action on Wilson lines is then,
\begin{equation}
    W_Q\to W_Q'=W_Q \exp\left(i\frac{Q}{\cN}\int \cN\Phi'\right)\ .
\end{equation}
We conclude that the action of the non-invertible defect on dyonic lines is:
\begin{equation}
    [(W_QH_P)^{\tilde r}]^k\to [(W_QH_P)^{\tilde r}]^k \exp\left(ik \frac{\tilde r Q}{\cN}\int \cN\Phi'\right)\exp\left(ik \frac{\tilde r P\cN}{\cM}\int \cM\Psi\right)\ .
\end{equation}
If we want the new operators to be trivial, we need 
\begin{equation}
    k=\text{lcm}\left(\frac{\cM}{\gcd(\cM,\tilde r P\cN)},\frac{\cN}{\gcd(\cN,\tilde r Q)}\right)=\frac{\cN\cM}{\gcd(\cM,\tilde r P)\gcd(\cN,\tilde r Q)}\ .
\end{equation}
This result is equivalent to \eqref{boundary condition non-invertible defect}. Another way to get this result is to realize that $(W_QH_P)^{\tilde r}$ has charge $q=\tilde r \gcd(\cM Q,\cN P)$ under the gauged $\bZ_{\cM\cN}\simeq \bZ^e_\cN \times \bZ_\cM^m$.\footnote{Indeed, under a $\bZ^e_\cN \times \bZ_\cM^m$ transformation of phases $(2\pi\frac{k_1}{\cN},2\pi\frac{k_2}{\cM})$, the operator $(W_Q H_P)^{\tilde r}$ acquires the phase $2\pi \frac{\tilde r}{\cN\cM} \left(\cM Q k_1 +\cN P k_2\right)$. For any choice of $k_1$ and $k_2$, we have $\cM Q k_1 +\cN P k_2=k_3 \gcd(\cM Q,\cN P)$ with $k_3\in \bZ$. The phase can therefore always be written as $2\pi \frac{k_3}{\cN\cM}\tilde r\gcd(\cM Q,\cN P)$.}
The operator $(W_QH_P)^{k \tilde r}$ therefore has a trivial charge if $k=\frac{\cM\cN}{\gcd(\cM\cN,q)}$.

Let us finally comment on the fact that a more general topological interface can be obtained by composing the two kinds that we have discussed in this section, namely the one that implements $SL(2,\bZ)$ transformations, and the one that implements a rescaling of the coupling (together with a change in the global boundary conditions). This composition also includes the generic gauging of $\bZ_\cN^e \times \bZ_\cM^m$ with a non-trivial discrete torsion. Together, these interfaces act on the bulk coupling $\tau$ as any general $SL(2,\bQ)$ transformation. We will show how to construct an explicit action for these more general interfaces using the SymTFT in Section \ref{setup}.

\section{Interacting Boundary Conditions and Boundary CFTs} \label{interacting bc}

Thus far, we have discussed free boundary conditions where no propagating edge mode is present. A larger subset of boundary conditions is obtained by considering a $U(1)$-symmetric 3d CFT coupled to the bulk through the action
\begin{equation}
    S_{\partial}=S_{CFT} + S_{free}(p,v,k) +\int_{\partial X_4} i(u^T\Phi) \wedge * J_{3d}\,,
\end{equation}
where $J_{3d}$ is the $U(1)$ current of the CFT and $u$ is some vector of integers.

When $k$ is non-degenerate, this new coupling modifies the boundary conditions to
\begin{equation}
    P\left(\frac{i}{e^2}*F-\frac{\theta}{2\pi}F\right)+QF= -2\pi P u^Tk^{-1}v*J_{3d}\,.
\end{equation}
Since $u$ is a vector of integers, the quantity $u^Tk^{-1}v$ has the same quantization as $k^{-1}v$. We then generically have $u^Tk^{-1}v=s\frac{g}{R}=s\frac{g}{rP}$, with $s\in \bZ$.  
In this case we get
\begin{equation}\label{eq: CFT b.c.}
    r \left( P\left(\frac{i}{e^2}*F-\frac{\theta}{2\pi}F\right)+QF\right)= -2\pi  sg *J_{3d}\,,
\end{equation}
as the new boundary condition which couples the 3d CFT to the bulk photon. This boundary condition can also be interpreted as defining a \emph{non-local} 3d CFT, which we denote by $B^{(p,v,k,u)}(\tau,\overline{\tau})$ \cite{DiPietro:2019hqe}. This theory arises from integrating out the bulk photon to obtain an effective three-dimensional description. However, due to interactions with the bulk, the resulting theory is generically non-local and lacks a stress tensor. A caveat is that these interactions may induce relevant deformations that spoil boundary conformal symmetry and produce an RG flow. As emphasized in \cite{DiPietro:2019hqe}, at the perturbative level one can always tune the boundary couplings to cancel such contributions and restore conformal invariance order by order in perturbation theory. However, in some cases non-perturbative effects may drive the boundary theory into a phase in which conformality is lost. If we wish to restrict to conformal boundary conditions, we must therefore assume that such effects are absent and exclude the 3d CFTs for which this occurs.

From \eqref{eq: CFT b.c.}, we conclude that the bulk 1-form symmetry operators, which were previously set to vanish at the boundary, now coincide with the $U(1)$ global symmetry of the CFT:
\begin{equation}\label{eq: CFT U(1)}
    \exp\left(i\frac{\alpha}{2\pi}\int  P\left(-\frac{i}{e^2}*F+\frac{\theta}{2\pi}F\right)-Q F\right)= \exp\left(i\frac{sg}{r}\alpha \int *J_{3d}\right)\,.
\end{equation}
This implies that bulk lines of the form $\left(W_N H_{M}\right)^n$,
with $N$ and $M$ satisfying $PN - QM = 1$ and $n \in \mathbb{Z}$, can now end at the boundary on local operators of the CFT with charge $q = n \frac{r}{sg}$. This ensures consistency with \eqref{eq: CFT U(1)}. Since the $U(1)$ charge $q$ must be quantized, only the bulk lines 
\begin{equation}
\left(W_N H_{M}\right)^{\frac{sg}{\gcd(sg,r)} n'} \quad n' \in \bZ
\end{equation}
are endable, terminating on local operators of charge 
\begin{equation}
q = n' \frac{r}{\gcd(sg,r)}\ .
\end{equation}
On the other hand, lines with $n \not \in \frac{sg}{\gcd(sg,r)} \bZ$ remain not endable on the boundary. However, boundary local operators with charge $q \not \in \frac{r}{\gcd(sg,r)}\bZ$ are still generically not gauge invariant since the boundary field $u^T \Phi$ plays the role of a dynamical gauge field for the boundary CFT. Therefore such local operators are still endpoints of the boundary line operators of $\Phi$.\footnote{Notice that the topological terms in the boundary action can trivialize certain line operators of $\Phi$. In such cases, a subset of the local operators may remain genuine local operators of the boundary theory.} 
Let us just comment that one can always choose $u$, and hence $s$, such that $\frac{r}{\gcd(sg,r)}=1$ and so all boundary local operators are attached to bulk lines, and the subtlety mentioned above does not arise. Such an overall rescaling of $u$ can be seen as a (boundary) discrete gauging.

When $k$ is degenerate, the boundary condition becomes
\begin{equation}
    v_0^Tv F= -2\pi v_0^Tu *J_{3d}\,.
\end{equation}
In particular, if we choose that $u=-v$ then,
\begin{equation}
    F= 2\pi *J_{3d}\,.
\end{equation}
This implies that the boundary global symmetry is identified with the bulk magnetic 1-form symmetry, namely
\begin{equation}
    U^{CFT}_{\alpha}:= \exp{\left(i\alpha \int * J_{3d}\right)} \equiv U^m_{\alpha}:= \exp{\left(i\alpha \int \frac{F}{2\pi}\right)}\,,
\end{equation}
and that the 't Hooft lines can end on the charged operators of the CFT.

Finally, the analysis of Section~\ref{sec: 3} can be readily extended to this case. The action of the topological interfaces is analogous to that in the free case, transforming the parameters $(p,v,k)$ while leaving $u$ invariant.

\subsection{Weak coupling limit and 3d CFTs}

These families of conformal boundary conditions $B^{(p,v,k,u)}(\tau, \overline{\tau})$ reduce to free boundary conditions times a decoupled \emph{local} CFT in the bulk weak coupling limit $e \to 0$~\cite{DiPietro:2019hqe}. To analyze this limit, as is customary in weakly coupled gauge theories, it is convenient to rescale the gauge field as $A = e A'$ in order to obtain a finite action in the limit. Using this rescaled variable and taking the limit $e \to 0$, we find that the boundary CFT decouples from the bulk, yielding:
\begin{equation}\label{eq: action 3d CFT}
    B^{(p,v,k,u)}(\tau, \overline \tau) \xrightarrow{e\rightarrow 0} S_{CFT} + i\int  (u^T \Phi) \wedge * J_{3d} +\frac{i}{4\pi}\int \Phi^T \wedge (kd\Phi) \,.
\end{equation}
As a result, we get that the weak coupling limit does not generate the local 3d CFT we started with, but a new theory. This can be interpreted as a refinement of the SL$(2,\mathbb{Z})$ action on three-dimensional CFTs, as discussed in~\cite{Witten:2003ya}. In addition to the standard $S$ and $T$ transformations, corresponding respectively to gauging a $U(1)$ symmetry without introducing a kinetic term (i.e. $u = 1$, $k = 0$) and stacking with a background Chern-Simons term (i.e. $u = 0$ and $k \in \bZ$), the action in~\eqref{eq: action 3d CFT} may also include more general topological operations, depending on the structure of the matrix $k$. These operations can involve, for example, gauging $\mathbb{Z}_N$ subgroups of the $U(1)$ global symmetry, possibly with discrete torsion $\xi\in H^3(B\bZ_N,U(1)) \cong \bZ_N$. A concrete example of this is given by choosing $u^T = (1,0)$ and $k = \begin{pmatrix}
\xi & N \\
N & 0
\end{pmatrix}$ for which the 3d local CFT becomes
\begin{equation}
    S_{3d} = S_{CFT} + i \int \Phi_1 \wedge * J_{3d} + \frac{iN}{2\pi}\int \Phi_1 \wedge d\Phi_2 + \frac{i \xi}{4\pi}\int \Phi_1 \wedge d \Phi_1\,. 
\end{equation}
This exactly corresponds to the $\bZ_N$-gauging of $S_{CFT}$ with discrete torsion $\xi$ \cite{Kapustin:2014gua}. 

Two standard examples are the Dirichlet- and Neumann-type boundary conditions coupled to the boundary CFT, which are described by the actions
\begin{align}
    S^D_{\partial} &= S_{\mathrm{CFT}} + \int \left( \frac{i}{2\pi} \Phi  \wedge dA - i \Phi \wedge * J_{3d} \right), \\
    S^N_{\partial} &= S_{\mathrm{CFT}} + \int \left( \frac{i}{2\pi} \Phi_1 \wedge d\Phi_2 + \frac{i}{2\pi} \Phi_2 \wedge  dA - i \Phi_1 \wedge * J_{3d} \right) = S_{\mathrm{CFT}}+ \int i A \wedge * J_{3d}\,,
\end{align}
where in the last equation we have integrated out $\Phi_2$.
In the weak coupling limit they generate the local CFTs
\begin{equation}
\begin{split}   
    &B^D(\tau, \overline{\tau}) \xrightarrow{e \to 0} S_{\mathrm{CFT}} - \int_{\partial X_4} i \Phi \wedge * J_{3d}\,,\\
    &B^N(\tau, \overline{\tau}) \xrightarrow{e \to 0} S_{\mathrm{CFT}}\ ,
\end{split}
\end{equation}
which correspond to the $S$-transform of the original 3d CFT, and just the original CFT, respectively.

As described in~\cite{DiPietro:2019hqe}, in addition to the decoupling point at $\tau = i\infty$, the bulk $SL(2,\mathbb{Z})$ duality implies the existence of additional points where the boundary CFT decouples from the bulk. The families of boundary conditions $B^{(p,v,k,u)}(\tau, \overline{\tau})$ exhibit such decoupling because, for a given limit value of $\tau$, there can exist duality frames in which the transformed gauge coupling $e$ vanishes. However, since an $SL(2,\mathbb{Z})$ transformation acts non-trivially on the boundary data $(p,v,k)$, the resulting local CFT at these additional decoupling points will in general take the form given in~\eqref{eq: action 3d CFT}. 

For instance, because bulk S-duality maps $\tau \rightarrow -\frac{1}{\tau}$ and it exchanges Dirichlet and Neumann boundary conditions, we find:
\begin{equation}
\begin{split}
    &B^D(\tau,\overline \tau)|_{\tau = 0} \overset{S}{\cong} B^N(\tau ,\overline \tau)|_{\tau = i\infty} \rightarrow S_{CFT}\,. \\
    &B^N(\tau,\overline \tau)|_{\tau = 0} \overset{S}{\cong} B^D(\tau ,\overline \tau)|_{\tau = i\infty} \rightarrow S_{CFT}+ \int_{\partial X_4} i \Phi \wedge * J_{3d} \,.
\end{split}
\end{equation}
It is easy to show that the generic decoupling point of any $B^{(p,v,k,u)}(\tau,\overline \tau)$ is at $\tau = -\frac{\mathsf{p}}{\mathsf{q}}$, for any $\mathsf{p},\mathsf{q} \in \bZ$ such that $\gcd(\mathsf{p},\mathsf{q})=1$ \cite{DiPietro:2019hqe}.

Finally, let us mention that at specific points on the bulk conformal manifold, there exist combinations of duality transformations and half-space gauging operations that leave the bulk theory invariant~\cite{Choi:2021kmx,Niro:2022ctq}. As extensively discussed in the literature, these combinations correspond to (non-invertible) global symmetries of the bulk theory. We can therefore study their action on the boundary conditions in the same spirit as the previous discussion on topological interfaces (see~\cite{Cordova:2023ent} for a related discussion). Since the bulk data remain invariant under these transformations, their actions imply equivalences between different boundary CFTs within the same bulk theory. In this direction, it would be also interesting to analyze the interplay between boundary conditions and 't Hooft anomalies of these non-invertible symmetries described in \cite{Antinucci:2023ezl,Cordova:2023bja}, generalizing the discussion of \cite{Choi:2023xjw}.

\section{The Boundary Symmetry TFT} \label{setup}
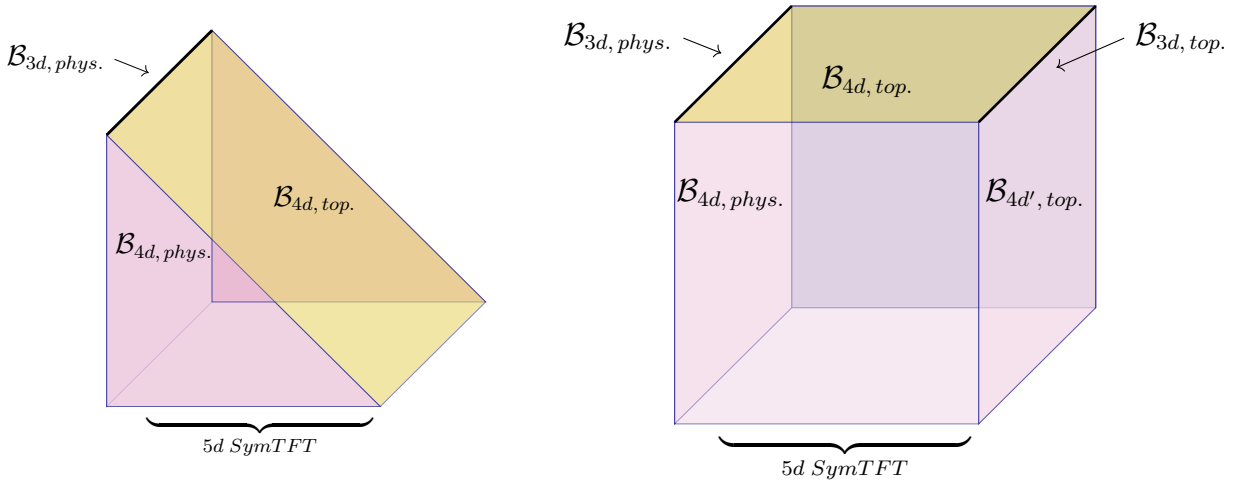
\begin{figure}[t]
      \centering
      \hfill
    \begin{minipage}{0.48\textwidth}
        \centering
        \newcommand{\DepthB}{4}
        \newcommand{\HeightB}{4}
        \newcommand{\WidthB}{4}
        \scalebox{0.9}{\begin{tikzpicture}
        \coordinate (O) at (0,0,0);
        \coordinate (A) at (0,\WidthB,0);
        \coordinate (B) at (\DepthB,0,0);
        \coordinate (C) at (0,0,\HeightB);
        \coordinate (D) at (0,\WidthB,\HeightB);
        \coordinate (E) at (\DepthB,0,\HeightB);

        \draw[blueflag,fill=mulberry!10,opacity=.8]  (O)--(A)--(D)--(C)--cycle;
        \draw[blueflag,fill=mulberry!30,opacity=.8]  (C)--(D)--(E)--cycle;
        \draw[blueflag,fill=mulberry!40,opacity=.8]  (O)--(A)--(B)--cycle;
        \draw[blueflag,fill=prettyyellow,opacity=.6]  (A)--(B)--(E)--(D)--cycle;

        \draw[line width=1.2,black] (A)--(D);

        \node at (-.7,0.8) {$\mathcal{B}_{4d, \,phys.}$};
        \node at (1.5,1.5) {$\mathcal{B}_{4d,\,top.}$};
        \node at (.7,-2) {$\underbrace{\qquad \qquad \qquad\qquad }_{5d \; SymTFT}$};

        \draw[->] (.5,5.5,5)--(1,5.3,5);
        \node[anchor=east] at (.5,5.5,5) {$\mathcal{B}_{3d,\, phys.}$};
        \end{tikzpicture}}
    \end{minipage}
  \begin{minipage}{0.48\textwidth}
        \centering
       \newcommand{\Depth}{4}
\newcommand{\Height}{4}
\newcommand{\Width}{4}
\scalebox{1}{\begin{tikzpicture}
\coordinate (O) at (0,0,0);
\coordinate (A) at (0,\Width,0);
\coordinate (B) at (0,\Width,\Height);
\coordinate (C) at (0,0,\Height);
\coordinate (D) at (\Depth,0,0);
\coordinate (E) at (\Depth,\Width,0);
\coordinate (F) at (\Depth,\Width,\Height);
\coordinate (G) at (\Depth,0,\Height);

\draw[blueflag,fill=blueflag!30] (O) -- (A) -- (E) -- (D) -- cycle;
\draw[blueflag,fill=mulberry!10] (O) -- (A) -- (B) -- (C) -- cycle;
\draw[blueflag,fill=mulberry!20,opacity=0.8] (D) -- (E) -- (F) -- (G) -- cycle;
\draw[blueflag,fill=mulberry!20,opacity=0.6] (C) -- (B) -- (F) -- (G) -- cycle;
\draw[blueflag,fill=prettyyellow, opacity=.6] (A) -- (B) -- (F) -- (E) -- cycle;

\node[] at (-.8,1.5) {$\mathcal{B}_{4d, \,phys.}$};
\node[] at (3.2,1.5) {$\mathcal{B}_{4d',\,top.}$};
\node at (.7,-2) {$\underbrace{\qquad \qquad \qquad\qquad }_{5d \; SymTFT}$};

\draw[line width = 1, black] (B) -- (A);
\draw[line width = 1, black] (F) -- (E);

\draw [->] (.5,5.5,5) -- (1,5.3,5);
\node[anchor=east] at (.5,5.5,5)  {$\mathcal{B}_{3d,\, phys.}$};
\draw [->] (6.3,5.5,5)--(5.6,5.3,5.5) ;
\node[anchor=west] at (6.3,5.5,5)  {$\mathcal{B}_{3d,\, top.}$};

    \node at (1,3.) {$\mathcal{B}_{4d ,\,top.}$};
     \end{tikzpicture}}
    \end{minipage}

    \caption{Two possible ways of constructing a Boundary SymTFT. Left: The physical 3d boundary serves as a corner between the non-topological and topological boundary conditions of the slab. Right: A new 3d topological corner is introduced as an interface between two 4d topological boundaries $\mathcal{B}_{4d ,top.}$ and $\mathcal{B}_{4d' ,top.}$. The choice of this new data determines the topological couplings of the physical 3d boundary condition.}
    \label{edges symtft}
\end{figure}

In this section, we aim to show how all the local and global properties of Maxwell boundary conditions are elegantly encoded in the Boundary Symmetry TFT description of the system, namely a five-dimensional TFT defined on a slab manifold with corners, see Figure \ref{edges symtft}. For details on the SymTFT with corners, we largely follow the setup and insights presented in~\cite{Bhardwaj:2024igy}.

In this context, we have two ways to define the boundary SymTFT geometry, both depicted in Figure \ref{edges symtft}. On the left-hand side of the figure, only one corner is present and it reproduces the 3d boundary condition upon compactification. A slightly more useful presentation is on the right-hand side. Indeed, we will show that, while the choice of the $\mathcal{B}_{4d',top}$ topological boundary condition determines the value of the bulk coupling $\tau'$ \cite{Paznokas:2025epc}, by spanning all the possible $\mathcal{B}_{4d ,top}$ topological boundary conditions and the corresponding $\mathcal{B}_{3d,top}$ topological interfaces, we are able to span the full set of free boundary conditions of Maxwell theory, determining both their local and global properties.

\begin{figure}[t]
    \centering
    \newcommand{\Depth}{4}
    \newcommand{\Height}{4}
    \newcommand{\Width}{4}

    \begin{tikzpicture}[scale=1]

    \begin{scope}
        \coordinate (O) at (0,0,0);
        \coordinate (A) at (0,\Width,0);
        \coordinate (B) at (0,\Width,\Height);
        \coordinate (C) at (0,0,\Height);
        \coordinate (D) at (\Depth,0,0);
        \coordinate (E) at (\Depth,\Width,0);
        \coordinate (F) at (\Depth,\Width,\Height);
        \coordinate (G) at (\Depth,0,\Height);

        \draw[blueflag,fill=mulberry,opacity=.8] (O) -- (C) -- (G) -- (D) -- cycle;
        \draw[blueflag,fill=blueflag!30] (O) -- (A) -- (E) -- (D) -- cycle;
        \draw[blueflag,fill=mulberry!10] (O) -- (A) -- (B) -- (C) -- cycle;
        \draw[blueflag,fill=mulberry!20,opacity=0.8] (D) -- (E) -- (F) -- (G) -- cycle;
        \draw[blueflag,fill=mulberry!20,opacity=0.6] (C) -- (B) -- (F) -- (G) -- cycle;
        \draw[blueflag,fill=prettyyellow, opacity=.6] (A) -- (B) -- (F) -- (E) -- cycle;

        \node[] at (-.8,1.5) {$\mathcal{B}_{4d, \,phys.}$};
        \node[] at (3.2,1.5) {$\mathcal{B}_{4d',\,top.}$};

        \draw[line width = 1, black] (B) -- (A);
        \draw[line width = 1, black] (F) -- (E);

        \draw [->] (.5,5.5,5) -- (1,5.3,5);
        \node[anchor=east] at (.5,5.5,5)  {$\mathcal{B}_{3d,\, phys.}$};
        \draw [->] (6.3,5.5,5)--(5.6,5.3,5.5) ;
        \node[anchor=west] at (6.3,5.5,5)  {$\mathcal{B}_{3d,\, top.}$};

        \node at (1,3.) {$\mathcal{B}_{4d ,\,top.}$};
    \end{scope}

    \draw[->, thick] (5.8,1.5) -- (8.2,1.5) node[midway, above] {\small oblique compactification};

    \begin{scope}[xshift=11cm]
        \coordinate (O) at (0,0,0);
        \coordinate (A) at (0,\Width,0);
        \coordinate (B) at (0,\Width,\Height);
        \coordinate (C) at (0,0,\Height);

        \coordinate (P1) at (0,\Width,\Height);           
        \coordinate (P2) at (0,\Width,0);                 
        \coordinate (P3) at (-1.535,-0.535+1);            
        \coordinate (P4) at (0,2);                        

        \draw[blueflag,fill=mulberry!10] (O) -- (A) -- (B) -- (C) -- cycle;

        \fill[prettyyellow, opacity=0.6] (P1) -- (P2) -- (P4) -- (P3) -- cycle;

        \node[] at (-.8,2.) {$\tau$};
        \node[] at (-.8,0.) {$\tau'$};

        \draw[line width = 1, black] (B) -- (A);
        \draw[line width = 1, black] (P3) -- (P4);

        \draw [->] (.5,5.5,5) -- (1,5.3,5);
        \node[anchor=east] at (.5,5.5,5)  {Boundary};
    \end{scope}

    \end{tikzpicture}

    \caption{Oblique slab compactification of the 5d theory on an interval. On the left, the full 5d/4d/3d setup with physical and topological boundaries. On the right, the phyiscal 4d theory with a topological interface between $\tau$ and $\tau'$ and a boundary. }
    \label{fig:interface}
\end{figure}
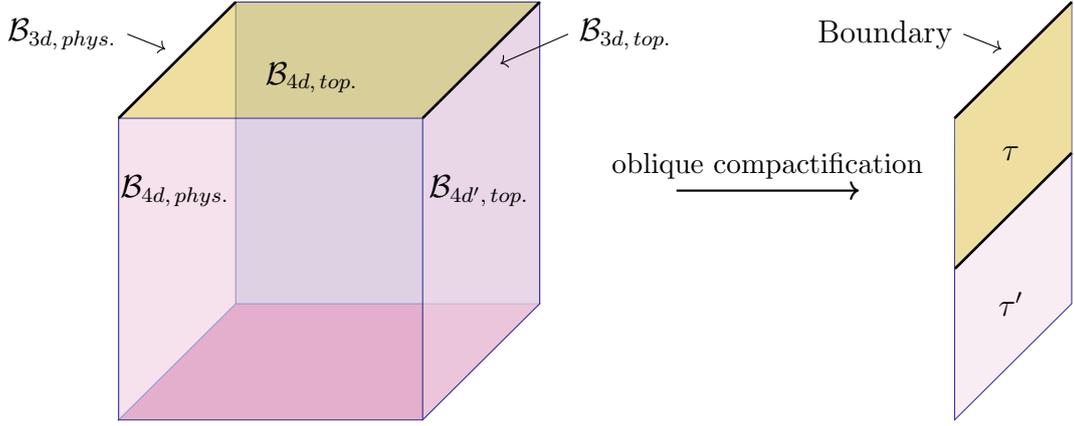

The Boundary SymTFT also provides a natural framework for implementing the various maps between boundary conditions, realized via the fusion of topological interfaces in Maxwell theory, as described in Section~\ref{sec: halfgauging int}. Indeed, rather than compactifying the five-dimensional slab geometry in the standard way, one can perform an \emph{oblique compactification}, leading to an intermediate situation in which a topological interface between two Maxwell theories appears in the four-dimensional physical theory (see Figure~\ref{fig:interface}).

\subsection{The 5d SymTFT}

We start by describing the bulk 5d TQFT which is given by the non-compact BF theory
\begin{equation} \label{5d bf theory}
    S_{5d} = \frac{i}{2\pi} \int_{M_5} b\wedge dc \,,
\end{equation}
which correctly captures the $U(1)^{(1)}_e \times U(1)^{(1)}_m$ continuous symmetries of the 4d Maxwell theory \cite{Antinucci:2024zjp,Brennan:2024fgj}. Here $b,c$ are real 2-form gauge fields, i.e. globally defined 2-forms on $M_5$.\footnote{Throughout this note, we adopt the convention that real gauge fields are denoted by lowercase letters, while $U(1)$ gauge fields are denoted by uppercase letters.} The gauge transformations of the fields $b$ and $c$ are:
\begin{equation}
    b \longmapsto b + d \lambda_b\quad \,, \quad c \longmapsto c+d\lambda_c\,,
\end{equation}
where $\lambda_b$ and $\lambda_c$ are globally defined differential forms ($\bR$ gauge fields do not admit large gauge transformations).
The equations of motion $db=0$ and $dc=0$ imply that the following gauge invariant surface operators are topological: 
\begin{equation}
    \begin{split}
        U_x(\Gamma) &= e^{ix \int_\Gamma b}\,, \qquad x \in \mathbb{R}\,, \\
        V_y(\Gamma) &= e^{iy\int_\Gamma c}\,, \qquad y \in \mathbb{R}\,,
    \end{split}
\end{equation}
for $\Gamma$ any 2-cycle in $M_5$. The fact that there is no quantization condition on the charges $x,y$ again follows from $b$ and $c$ being $\mathbb{R}$ gauge fields. 

The canonical quantization of this BF theory implies that the linking between surface operators is antisymmetric:
\begin{equation}
    \langle U_xV_y (\Gamma) \ U_{x'} V_{y'}(\Gamma') \rangle = e^{2\pi i (xy'-yx')\mathrm{Link}(\Gamma, \Gamma')} \,.
\end{equation}

\subsubsection{The \texorpdfstring{$\mathcal{B}_{4d,top.}$}{B4d} topological and \texorpdfstring{$\mathcal{B}_{4d,phys.}$}{B4d} physical boundary conditions} \label{4d section}

Let us now consider adding two boundaries to this 5d TFT, as shown on the left of Figure \ref{edges symtft}. Gauge invariance then requires imposing appropriate conditions on $\mathcal{B}_{4d,top.}$. In practice, this is done via introducing the edge mode $A$ which is a $U(1)$ 1-form gauge field. We start with the most general topological boundary condition for a theory with $U(1)\times U(1)$ symmetry, parametrized by choices of constants $p,q,m,n$ such that $pn-qm=1$:
\begin{equation} \label{4d general}
    S_{4d, \text{top.}} = \frac{i}{2\pi}\int_{M_4} dA\wedge(pc+qb)+\frac{mp}{2}c\wedge c+qm b \wedge c+\frac{qn}{2}b\wedge b\,.
\end{equation}
Note that since $b$ and $c$ are $\bR$ gauge fields, the coefficients $p,q,m,n$ need not be integers. Under gauge transformations, we require that
\begin{equation}
    A \longmapsto A + d \lambda_A - m \lambda_c - n \lambda_b\,.
\end{equation}
Together with the condition $pn-qm=1$, this ensures gauge invariance of the combined action $S_{5d} + S_{4d,\text{top.}}$. 

The equations of motion obtained when varying with respect to $c$ and $b$ respectively are
\begin{equation}
    pdA+mpc+(qm+1)b=0\,, \quad qdA+qmc+qnb=0\,.
\end{equation}
The consistency of these e.o.m. follows from the condition $pn-qm=1$. The sum over fluxes of $dA$ also implies the following boundary constraint:
\begin{equation}
\begin{split}
\int pc+qb =: \int d\tilde{A} \in 2\pi \bZ\,,
\end{split}
\end{equation}
where $\tilde A$ is a new $U(1)$ gauge field.
Inverting these relations leads to:
\begin{equation}
\begin{split}
b&=-md\tilde{A}-pdA\,,\\
c&=nd\tilde{A}+qdA\,.
\end{split}
\end{equation}
Due to the above choice of boundary conditions, this implies that the following surface operators are able to end on the 4d topological boundary:
\begin{equation}
    (U_nV_m)^\ell \quad , \quad (U_qV_p)^s \quad \quad \ell,s\in \bZ\,,
\end{equation}
and they form the Lagrangian algebra for this topological boundary.\footnote{In this context, a Lagrangian algebra is a non-simple topological surface operator constructed as a direct sum of the maximal subset of commuting surfaces $U_xV_y$.}

This class of boundary conditions captures all possible $SL(2,\bR)$ transformations of the Maxwell topological boundary. To see this, let us consider the symmetrized version of the 5d bulk action:
\begin{equation}
    \frac{i}{4\pi}\int b\wedge dc-c\wedge db\,,
\end{equation}
which differs from \eqref{5d bf theory} by a boundary term and which is invariant under $SL(2,\bR)$ acting on the bulk gauge fields as
\begin{equation} \label{sl on bulk fields}
    \begin{pmatrix}
        c\\
        b
    \end{pmatrix}\longmapsto \begin{pmatrix}
        x & y\\
        w & z
    \end{pmatrix}\begin{pmatrix}
        c\\
        b
    \end{pmatrix} \, ,\quad \quad \begin{pmatrix}
         x & y\\
        w & z
    \end{pmatrix}\in SL(2,\bR)\,.
\end{equation}
Writing the 4d topological boundary action as\footnote{Note that the appearance of the $\frac{1}{2}b\wedge c$ term, both in $S_{4d,\text{top.}}$ here and in $S_{4d,\text{phys.}}$ below, is due to writing the bulk action in the symmetrized presentation.}
\begin{equation}
    \begin{split}
        S_{4d,\text{top.}}&=\frac{i}{2\pi}\int dA\wedge (pc+qb)+\frac{mp}{2}c \wedge c+qm b\wedge c+\frac{qn}{2}b\wedge b+\frac{1}{2}b\wedge c\\
        &=\frac{i}{2\pi}\int dA \wedge (pc+qb)+\frac{1}{2}(pc+qb)\wedge (mc+nb)\,,
    \end{split}
\end{equation}
again with $p,q,m,n\in \bR$ satisfying $pn-qm=1$, makes it clear that $p,q,m,n\in\bR$ parameterize the $SL(2,\bR)$-orbit of boundary conditions containing the standard Dirichlet one. Indeed, an $SL(2,\bR)$-rotation of the bulk fields $b$ and $c$ as in \cref{sl on bulk fields} amounts to the following $SL(2,\bR)$-action on the parameters $p,q,m,n$:
\begin{equation}\label{SL2R transformation}
    \begin{pmatrix}
        p & q\\
        m & n
    \end{pmatrix}\longmapsto\begin{pmatrix}
        p & q\\
        m & n
    \end{pmatrix}\begin{pmatrix}
         x & y\\
        w & z
    \end{pmatrix}\,.
\end{equation}
On the 4d physical boundary we consider the action\footnote{Boundary conditions structurally analogous to $S_{4d,\text{phys.}}$ have been previously examined in \cite{Maldacena:2001ss}, where they were shown to give rise to generalized Maxwell theories. See also \cite{Benini:2022hzx, Antinucci:2024bcm,Argurio:2024ewp}.}
\begin{equation}
    S_{4d, \text{phys.}} = \frac{1}{4\pi}\int (-c\wedge *c+i b\wedge c)\,.
\end{equation}
After slab compactification, we are left with the following 4-dimensional action:\footnote{For slab compactification the boundaries are oriented; we consider the $4d$ action $S_{4d,\text{top.}}-S_{4d,\text{phys.}}$.}
\begin{equation}
    S_{4d}= \frac{1}{4\pi}\int c\wedge *c+\frac{i}{2\pi}\int dA\wedge  (pc+qb)+\frac{mp}{2}c \wedge c+qm b \wedge c+\frac{qn}{2}b\wedge b\,,
\end{equation}
whose equations of motion yield
\begin{equation}
    i*c -(pdA +mpc +qmb)=0 \, , \quad q(dA+mc+nb)=0\,,
\end{equation}
such that 
\begin{equation}
    \begin{split}
        c&=\frac{-m}{m^2+n^2}dA-\frac{in}{m^2+n^2}*dA\,, \\
        b&=-\frac{n}{m^2+n^2}dA+\frac{im}{m^2+n^2}*dA \,.
    \end{split}
\end{equation}
Inserting these relations back into $S_{4d}$, one obtains \begin{equation}
    S_{4d}= \frac{1}{4\pi (m^2+n^2)}\int dA\wedge *dA-\frac{i}{4\pi}\frac{pm+qn}{m^2+n^2}\int dA\wedge dA\,,
\end{equation}
which is the action of 4d Maxwell theory with non-zero $\theta$ angle, with the identifications
\begin{equation}
    e^2=m^2+n^2 \,, \quad -\frac{\theta}{2\pi}=\frac{pm+qn}{m^2+n^2}\,.
\end{equation}
\begin{figure}
    \centering
    \scalebox{.6}{
\begin{tikzpicture}
\node at (-2,4) {$a)$};

\coordinate (O) at (0,0,0);
\coordinate (A) at (0,4,0);
\coordinate (B) at (0,4,4);
\coordinate (C) at (0,0,4);
\coordinate (D) at (4,0,0);
\coordinate (E) at (4,4,0);
\coordinate (F) at (4,4,4);
\coordinate (G) at (4,0,4);

\draw[blueflag,fill=mulberry,opacity=.8] (O) -- (C) -- (G) -- (D) -- cycle;
\draw[blueflag,fill=blueflag!30] (O) -- (A) -- (E) -- (D) -- cycle;
\draw[blueflag,fill=mulberry!10] (O) -- (A) -- (B) -- (C) -- cycle;
\draw[blueflag,fill=mulberry!20,opacity=0.8] (D) -- (E) -- (F) -- (G) -- cycle;
\draw[blueflag,fill=mulberry!20,opacity=0.6] (C) -- (B) -- (F) -- (G) -- cycle;
\draw[blueflag,fill=prettyyellow, opacity=.6] (A) -- (B) -- (F) -- (E) -- cycle;

\node[] at (-.8,1.5) {$\mathcal{B}_{4d\,phys.}$};
\node[] at (3.2,1.5) {$\mathcal{B}_{4d',top.}$};
\node[] at (1.3,3.2) {$\mathcal{B}_{4d,top.}$};

   \begin{scope}[shift={(-.5, 0.)}]

            \draw[blueflag,fill=pink,opacity=0.4] (4, 0,0) -- (4, 4, 0) -- (4, 4, 4) -- (4, 0, 4) -- cycle;
            \node at (3.25,1) {$\mathcal{C}$};

	\end{scope}

\node at (5,1) {$=$};
\begin{scope}[shift={(8,0)}]

\coordinate (O) at (0,0,0);
\coordinate (A) at (0,4,0);
\coordinate (B) at (0,4,4);
\coordinate (C) at (0,0,4);
\coordinate (D) at (4,0,0);
\coordinate (E) at (4,4,0);
\coordinate (F) at (4,4,4);
\coordinate (G) at (4,0,4);

\draw[blueflag,fill=mulberry,opacity=.8] (O) -- (C) -- (G) -- (D) -- cycle;
\draw[blueflag,fill=blueflag!30] (O) -- (A) -- (E) -- (D) -- cycle;
\draw[blueflag,fill=mulberry!10] (O) -- (A) -- (B) -- (C) -- cycle;
\draw[blueflag,fill=mulberry!20,opacity=0.8] (D) -- (E) -- (F) -- (G) -- cycle;
\draw[blueflag,fill=mulberry!20,opacity=0.6] (C) -- (B) -- (F) -- (G) -- cycle;
\draw[blueflag,fill=prettyyellow, opacity=.6] (A) -- (B) -- (F) -- (E) -- cycle;

\node[] at (-.8,1.5) {$\mathcal{B}_{4d\,phys.}$};
\node[] at (3.2,1.5) {$\tilde{\mathcal{B}}_{4d',top.}$};
\node[] at (1.3,3.2) {$\mathcal{B}_{4d,top.}$};
\draw [->] (6.3,5.5,5)--(5.6,5.3,5.5) ;
\node[anchor=west] at (6.3,5.5,5)  {$\tilde{\mathcal{B}}_{3d,\, top.}$};
    
\end{scope}

\begin{scope}[shift={(0,-7)}]
\node at (-2,4) {$b)$};

\coordinate (O) at (0,0,0);
\coordinate (A) at (0,4,0);
\coordinate (B) at (0,4,4);
\coordinate (C) at (0,0,4);
\coordinate (D) at (4,0,0);
\coordinate (E) at (4,4,0);
\coordinate (F) at (4,4,4);
\coordinate (G) at (4,0,4);

\draw[blueflag,fill=mulberry,opacity=.8] (O) -- (C) -- (G) -- (D) -- cycle;
\draw[blueflag,fill=blueflag!30] (O) -- (A) -- (E) -- (D) -- cycle;
\draw[blueflag,fill=mulberry!10] (O) -- (A) -- (B) -- (C) -- cycle;
\draw[blueflag,fill=mulberry!20,opacity=0.8] (D) -- (E) -- (F) -- (G) -- cycle;
\draw[blueflag,fill=mulberry!20,opacity=0.6] (C) -- (B) -- (F) -- (G) -- cycle;
\draw[blueflag,fill=prettyyellow, opacity=.6] (A) -- (B) -- (F) -- (E) -- cycle;

\node[] at (-.8,1.5) {$\mathcal{B}_{4d\,phys.}$};
\node[] at (3.2,1.5) {$\mathcal{B}_{4d',top.}$};
\node[] at (1.3,3.2) {$\mathcal{B}_{4d,top.}$};

   \begin{scope}[shift={(0, -.5)}]

            \draw[blueflag,fill=pink,opacity=0.4] (0, 4,0) -- (0, 4, 4) -- (4, 4, 4) -- (4, 4, 0) -- cycle;
            \node at (1,3.25) {$\mathcal{C}$};

	\end{scope}

\node at (5,1) {$=$};
\begin{scope}[shift={(8,0)}]

\coordinate (O) at (0,0,0);
\coordinate (A) at (0,4,0);
\coordinate (B) at (0,4,4);
\coordinate (C) at (0,0,4);
\coordinate (D) at (4,0,0);
\coordinate (E) at (4,4,0);
\coordinate (F) at (4,4,4);
\coordinate (G) at (4,0,4);

\draw[blueflag,fill=mulberry,opacity=.8] (O) -- (C) -- (G) -- (D) -- cycle;
\draw[blueflag,fill=blueflag!30] (O) -- (A) -- (E) -- (D) -- cycle;
\draw[blueflag,fill=mulberry!10] (O) -- (A) -- (B) -- (C) -- cycle;
\draw[blueflag,fill=mulberry!20,opacity=0.8] (D) -- (E) -- (F) -- (G) -- cycle;
\draw[blueflag,fill=mulberry!20,opacity=0.6] (C) -- (B) -- (F) -- (G) -- cycle;
\draw[blueflag,fill=prettyyellow, opacity=.6] (A) -- (B) -- (F) -- (E) -- cycle;

\node[] at (-.8,1.5) {$\mathcal{B}_{4d\,phys.}$};
\node[] at (3.2,1.5) {$\mathcal{B}_{4d',top.}$};
\node[] at (1.3,3.2) {$\tilde{\mathcal{B}}_{4d,top.}$};
\draw [->] (6.3,5.5,5)--(5.6,5.3,5.5) ;
\node[anchor=west] at (6.3,5.5,5)  {$ \tilde{\mathcal{B}}_{3d,\, top.}$};
    
\end{scope}
    
\end{scope}

\end{tikzpicture}} 
    \caption{Illustration of the SL$(2,\mathbb{Z})$ 0-form symmetry, realized as a codimension-one defect of the 5d TFT, acting on the topological boundaries. In (a) it transforms the $4d'$ boundary, shifting $\tau'$ and leaving $4d$ fixed; in (b) it acts on $4d$, leaving $4d'$ unchanged. In both cases, it induces an action on the 3d topological corner.}
    \label{defect placement}
\end{figure}
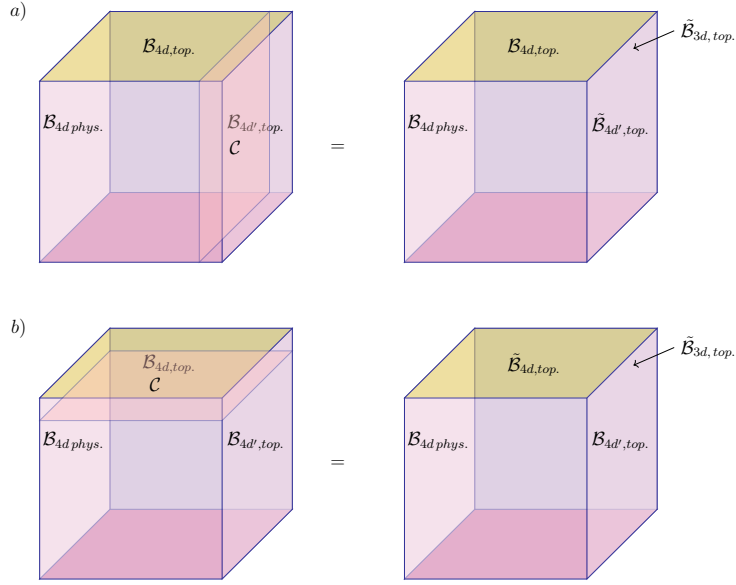
Equivalently 
\begin{equation} \label{86}
    \tau=\frac{\theta}{2\pi}+\frac{i}{e^2}=\frac{-pm-qn+i}{m^2+n^2}= \frac{pi-q}{-mi+n}\,.
\end{equation}
This is consistent with the fact that any point in the conformal manifold can be reached by an $SL(2,\bR)$ transformation (for instance starting from $\tau=i$, as in the present case).

\subsection{The \texorpdfstring{$SL(2,\bZ)$}{SL(2,Z)} duality from the SymTFT}

The identification of $\tau$ in \eqref{86} allows us to deduce the action of $SL(2,\bZ)$ transformations on the parameters $p$, $q$, $m$ and $n$.
Their action on $\tau$ 
\begin{equation} \label{144}
    \tau\to \frac{\mathcal{A}\tau+\mathcal{B}}{\mathcal{C}\tau+\mathcal{D}}\ ,
\end{equation}
translates to
\begin{equation}
\begin{pmatrix}
        p & q\\
        m & n
    \end{pmatrix} \to \begin{pmatrix}
        \mathcal{A} & -\mathcal{B}\\
        -\mathcal{C} & \mathcal{D}
    \end{pmatrix}\begin{pmatrix}
        p & q\\
        m & n
    \end{pmatrix}\,  .
\end{equation}
These transformations generate a $SL(2,\bZ)$ subgroup of the bulk $SL(2,\bR)$ symmetry. Since $SL(2,\bR)$ acts on the boundary parameters through \eqref{SL2R transformation}, the $SL(2,\bZ)$ subgroup takes the following form:
\begin{equation}\label{eq: SL2z 5d}
\mathfrak{M}=\begin{pmatrix}
        p & q\\
        m & n
    \end{pmatrix}^{-1}\begin{pmatrix}
        \mathcal{A} & -\mathcal{B}\\
        -\mathcal{C} & \mathcal{D}
    \end{pmatrix}\begin{pmatrix}
        p & q\\
        m & n
    \end{pmatrix}\, ,\quad \mathfrak{M} \in SL(2,\bZ)\, .
\end{equation}
These $SL(2,\bZ)$ transformations correspond to the subgroup of the bulk $SL(2,\bR)$ symmetry that does not alter the Lagrangian algebra trivialized on the topological boundary. As such, these matrices depend explicitly on the boundary parameters.
Two topological boundary conditions related by an $SL(2,\bZ)$ transformation are therefore equivalent in the sense that they will produce the same operator content in the physical theory.

Altering the topological boundary can be interpreted as performing a topological manipulation in the physical theory, such as gauging a discrete symmetry or a continuous symmetry with flat connections. This is made explicit in the SymTFT by noting that a generic element of the $SL(2,\bR)$ global symmetry corresponds to a condensation defect obtained by higher-gauging a subgroup of the $\bR \times \bR$ 2-form symmetry of the bulk BF theory \cite{Paznokas:2025epc}. Upon slab compactification, a condensation defect placed parallel to the boundary implements a specific flat gauging, determined by the choice of Lagrangian algebra for the topological boundary condition (see Figure \ref{defect placement}-a). In particular, the action of an $SL(2,\bZ)$ transformation does not modify the Lagrangian algebra and, therefore, the associated topological manipulation should be trivial. This is indeed what was observed in Section \ref{sec: halfgauging int}, where it was shown that any $SL(2,\bZ)$ transformation can be accomplished by gauging a $\bZ_1\times \bZ_1$ symmetry.

The parameters $e$ and $\theta$ are invariant under the following $SO(2)$ transformations,
\begin{equation}
    \begin{pmatrix}
       p & q\\
        m & n\\
    \end{pmatrix}\rightarrow  \begin{pmatrix}
       p & q\\
        m & n\\
    \end{pmatrix}\begin{pmatrix}
        \cos \sigma & \sin \sigma\\
        -\sin \sigma & \cos \sigma
    \end{pmatrix}\,.
\end{equation}
Indeed, it has been shown in \cite{Paznokas:2025epc} that $SO(2)\subset SL(2,\bR)$ correspond to (possibly non-invertible) symmetries of Maxwell theory. It is important to note, however, that these transformations do not preserve the topological boundary on its own. We are able to utilize such an $SO(2)$ transformation in order to set $m=0$ in $S_{4d}$ without changing the physical theory and without loss of generality. 

To illustrate more clearly what has been discussed thus far, namely that this $4d$ boundary reproduces Maxwell with $\tau = \frac{i}{e^2}+\frac{\theta}{2\pi}$ (as in \cite{Paznokas:2025epc}), we can choose the specific values $p=\frac{1}{e}$, $q=-\frac{\theta e}{2\pi}$, $n=e$, $m=0$:
\begin{equation} \label{4dtop}
    S_{4d,\mathrm{top.}}= \frac{i}{2\pi} \int_{M_4} dA\wedge\left(\frac{1}{e}c - \frac{e\theta}{2\pi} b \right) - \frac{ie^2\theta}{8\pi^2}\int_{M_4} b\wedge b\,.
\end{equation}
The gauge transformations of $A$ become,
\begin{equation}
    A \to A- e\lambda_b+d\lambda_A\,.
\end{equation}
With this choice of coefficients, the following bulk surface operators are trivially endable on $S_{4d}$,
\begin{equation}
    \begin{split}
        U_{ke}(\Gamma) &= e^{ike\int_\Gamma b} = e^{-ik\int_\gamma A}\,,\\
        U_{-\frac{e\theta l}{2\pi}} V_{\frac le} (\Gamma) &=e^{-i\frac{e\theta l}{2\pi}\int_\Gamma b} \; e^{i\frac le\int_\Gamma c} = e^{il\int_\gamma \tilde{A}}\,,
    \end{split}
\end{equation}
where $\partial \Gamma =\gamma$ and $k,l\in\bZ$.
The symmetry generators are then the remaining, non-trivialized surface operators with charges in $ \bR/ \bZ$ such that they generate a $U(1)^e \times U(1)^m$ 1-form symmetry. 

To summarize, we conclude that after slab compactification, combining topological and physical boundaries and evaluating everything on-shell, the result is 4d Maxwell with a theta term,
\begin{equation}
    S_{Maxwell}= \frac{1}{4\pi e^2}\int_{M_4}  F\wedge * F + \frac{i\theta}{8\pi^2}\int_{M_4}  F \wedge F\,,
\end{equation}
described by the complex coupling $\tau= \frac{i}{e^2}+ \frac{\theta}{2\pi}$.

\subsection{The \texorpdfstring{$\mathcal{B}_{4d' ,top.}$}{B4d} topological boundary condition}
Previously, we only considered the `wedge setup', see the left-hand side of Figure \ref{edges symtft}, wherein the 4d physical and topological boundaries were only separated by a 3d physical boundary. Now, we can move on to consider a `box setup', where there also exists a 3d topological boundary which separates two 4d topological boundaries which we name $4d$ and $4d'$ respectively. This is the setup pictured on the right-hand side of Figure \ref{edges symtft}. In this case, the $4d$ topological boundary interpolates between the three-dimensional physical and topological boundaries. The $4d'$ topological boundary corresponds to a $U(1)$ Maxwell theory with $\tau$ transformed to $\tau'$.
 
To this end, we employ the same form of the action as in Eq.~\eqref{4d general}, but we distinguish it in this context by using primes on the parameters and edge modes:
\begin{equation} \label{253}
    S_{4d' } = \frac{i}{2\pi} \int_{4d' } dA' \wedge (p'c +q'b) + \frac{m'p'}{2} c \wedge c + q'm' b \wedge c +\frac{q'n'}{2} b \wedge b\,.
\end{equation}
Again, we require that $p'n'-q'm'=1$ for gauge invariance and consistent e.o.m. The gauge transformations of $A'$ are
\begin{equation}
    A' \to A' + d \lambda_{A'} - m' \lambda_c - n' \lambda_b\,.
\end{equation}

As the $4d' $ topological boundary takes the same form as on $4d$, only with potentially different values of the parameters, this implies that upon \textit{oblique compactification} the result is an interface between two Maxwell theories, one with $\tau$ and the other $\tau'$, see Figure \ref{fig:interface}. We shall comment in the following on the possible limitations of having an interface which separates $\tau$ from any arbitrary $\tau'$.

\subsection{The \texorpdfstring{$\mathcal{B}_{3d}$}{B3d} topological interface}\label{sec: B3d topological interface}

The last remaining ingredients we must specify for our setup are the two 3d corners. Let us start by describing the 3d topological interface 
separating the $4d$ boundary from the $4d'$ one. We will fix the $3d$ topological interface in such a way to ensure that the $4d+4d' +3d$ system is gauge invariant. 

To go about constructing a gauge invariant interface, it is useful to fold the $4d' $ boundary on top of the $4d$ theory. After folding, $b$ and $c$ are now only defined on the 4d manifold and the action of the folded theory is
\begin{equation}\label{folded 4d}
    \begin{split}
        S_{4d}-S_{4d' }=&\frac{i}{2\pi}\int \begin{pmatrix}
        dA & dA'
    \end{pmatrix}\begin{pmatrix}
        p & q\\
        -p' & -q'
    \end{pmatrix}\begin{pmatrix}
        c\\ b
    \end{pmatrix} \\
    &+\frac{1}{2} \begin{pmatrix}
        c & b
    \end{pmatrix}\begin{pmatrix}
        pm-p'm' & pn-p'n'\\
        pn-p'n' & qn-q'n'
    \end{pmatrix}\begin{pmatrix}
        c \\ b
    \end{pmatrix}\ .
    \end{split}
\end{equation}
Note that the gauge transformations of $A$ and $A'$ can be summarized as 
\begin{equation}
    \begin{pmatrix}
        A \\
        A'
    \end{pmatrix} \rightarrow \begin{pmatrix}
        A \\
        A'
    \end{pmatrix}-\begin{pmatrix}
        m & n\\
        m' & n'
    \end{pmatrix} \begin{pmatrix}
        \lambda_c \\
        \lambda_b
    \end{pmatrix}\ .
\end{equation}
We then observe that (using $pn-p'n'=qm-q'm'$)
\begin{equation}
    \begin{pmatrix}
        pm-p'm' & pn-p'n'\\
        pn-p'n' & qn-q'n'
    \end{pmatrix}
    =\begin{pmatrix}
        p & q\\
        -p' & -q'
    \end{pmatrix}^T\begin{pmatrix}
        m & n\\
        m' & n'
    \end{pmatrix} \ .
\end{equation}
It is then easy to show that in the gauge variation of \eqref{folded 4d} all the terms where the fields $c$ and $b$ appear cancel, leaving only the terms linear in $dA$ and $dA'$ which are total derivatives.

The 3d interface between $4d$ and $4d'$ can now be interpreted as a (topological) boundary of the action \eqref{folded 4d}. To be well-defined, this boundary should cancel all the gauge transformations. To find a boundary satisfying this constraint, we can introduce $U(1)$ edge modes for $b$ and $c$, dubbed $\Phi$, $\Psi$, with gauge transformations
\begin{equation}
     \begin{pmatrix}
        d\Phi \\
        d\Psi
    \end{pmatrix} \rightarrow \begin{pmatrix}
        d\Phi \\
        d\Psi
    \end{pmatrix}+\mathcal{D}^{-1} \begin{pmatrix}
        \lambda_c \\
        \lambda_b
    \end{pmatrix}
\end{equation}
where $\mathcal{D}\in GL(2,\bR)$.\footnote{One can also consider $\bR$-valued edge modes. However, this will produce a boundary condition for the physical theory equipped with these non-compact degrees of freedom, which we want to avoid. We comment on these more generic boundary conditions in the next Section.} A generic gauge invariant topological boundary can now be written as\footnote{This form of the 3d interface is, in general, not closed under fusion. However, as we will show, the 3d interface in this form reproduces the most general local and global aspects of the boundary conditions, meaning we simply have freedom in how we can describe the same boundary conditions using different TQFTs.}
\begin{equation}
    \begin{split}
        S_{3d}&=\frac{-i}{2\pi}\int \begin{pmatrix}
        dA & dA'
    \end{pmatrix}\begin{pmatrix}
        p & q\\
        -p' & -q'
    \end{pmatrix}\mathcal{D}\begin{pmatrix}
        \Phi\\ \Psi
    \end{pmatrix}\\
    &+\frac{1}{2} \begin{pmatrix}
        \Phi & \Psi
    \end{pmatrix}\mathcal{D}^T\begin{pmatrix}
        pm-p'm' & pn-p'n'\\
        pn-p'n' & qn-q'n'
    \end{pmatrix}\, \mathcal{D}d\begin{pmatrix}
        \Phi \\ \Psi
    \end{pmatrix}\,.
    \end{split}
\end{equation}
Since all fields are $U(1)$ fields, this action is only well-defined if all entries in the matrices
$\begin{pmatrix}
        p & q\\
        -p' & -q'\end{pmatrix}\mathcal{D}$ and $\mathcal{D}^T\begin{pmatrix}
        pm-p'm' & pn-p'n'\\
        pn-p'n' & qn-q'n'
\end{pmatrix}\mathcal{D}$ are integers. 
For convenience, we can introduce the following notation
\begin{equation}
    \begin{pmatrix}
        x & y\\
        w & z
    \end{pmatrix}:=\begin{pmatrix}
       n'p-m'q & p'q-q'p\\
       n'm-m'n & p'n-q'm
    \end{pmatrix}=\begin{pmatrix}
        p & q\\
        m & n
    \end{pmatrix}\begin{pmatrix}
        p' & q'\\
        m' & n'
    \end{pmatrix}^{-1}\, ,
\end{equation}
which satisfies $xz-yw=1$. In particular, the elements $x,y,w,z$ are the parameters of an $SL(2,\bR)$ transformation that maps $p',q',m',n'\to p,q,m,n$ as
\begin{equation}
        \begin{pmatrix}
        p & q\\
        m & n
    \end{pmatrix}=\begin{pmatrix}
        x & y\\
        w & z
    \end{pmatrix}\begin{pmatrix}
        p' & q'\\
        m' & n'
    \end{pmatrix}\ .
\end{equation}
It is easy to see that a choice of $\cD$ exists if and only if $\begin{pmatrix}
        x & y\\
        w & z
    \end{pmatrix} \in SL(2,\bQ)$,
    meaning that we can describe interfaces separating two Maxwell theories related by an $SL(2,\bQ)$ transformation.
    
Let us start by considering the case where $\begin{pmatrix}
        x & y\\
        w & z
    \end{pmatrix}\in SL(2,\bZ)$. In this case, we can take \begin{equation}
        \mathcal{D}=\begin{pmatrix}
    p & q\\
    m & n
\end{pmatrix}^{-1}\, ,
    \end{equation} which gives
\begin{equation}
    S_{3d}=-\frac{i}{2\pi}\int \begin{pmatrix}
        dA & dA'
    \end{pmatrix}\begin{pmatrix}
        1 & 0\\
        -z & y
    \end{pmatrix}\begin{pmatrix}
        \Phi\\ \Psi
    \end{pmatrix}+\frac{1}{2} \begin{pmatrix}
        \Phi & \Psi
    \end{pmatrix}\begin{pmatrix}
      wz  & -wy\\
       -wy & xy
    \end{pmatrix}d\begin{pmatrix}
        \Phi \\ \Psi
    \end{pmatrix}\,.
\end{equation}
Note that we could try to integrate out the edge modes from this action. We obtain straightforwardly
\begin{equation}
    d\begin{pmatrix}
        \Phi \\ \Psi
    \end{pmatrix}=-\begin{pmatrix}
        \frac x w & -\frac 1 w\\
        1 & 0
    \end{pmatrix}d \begin{pmatrix}
        A \\ A'
    \end{pmatrix}\ .
\end{equation}
In general, we cannot substitute $\Phi$ and $\Psi$ for $A$ and $A'$ while preserving the correct quantization of the various $U(1)$ gauge fields. However, we see that it is indeed possible to make the single substitution $\Psi=-A$, thus eliminating one of the edge modes while keeping gauge invariance of the 4$d$-3$d$-4$d'$ system. The interface action that we get is
\begin{equation}
    S_{3d}=\frac{i}{2\pi}\int yA'\wedge dA+z A'\wedge d\Phi -\frac{xy}{2}A\wedge dA-xzA\wedge d\Phi-\frac{wz}{2}\Phi\wedge  d\Phi\,.
\end{equation}
One can easily see that this is the same form as the interface action \eqref{3d 2edge modes}, upon the identification 
\begin{equation}\label{SL2Z identification}
    \begin{pmatrix}
        \alpha & \beta\\
        \gamma & \delta
    \end{pmatrix}=-
    \begin{pmatrix}
        x & y\\
        w & z
    \end{pmatrix}^{-1}\ .
\end{equation}
When $\begin{pmatrix}
        x & y\\
        w & z
    \end{pmatrix}\in SL(2,\mathbb{Q})$, we can write
    \begin{equation}\label{SL2Q parametrization}
        \begin{pmatrix}
        x & y\\
        w & z
    \end{pmatrix}=\begin{pmatrix}
        \frac{x_n}{x_d} & \frac{y_n}{y_d}\\
        \frac{w_n}{w_d} & \frac{z_n}{z_d}
    \end{pmatrix}
    \end{equation}
where all parameters on the right-hand side are integers. In this case, we can consider
\begin{equation}\label{eq: D sl2q}
    \mathcal{D}=\begin{pmatrix}
        p & q\\
        m & n
    \end{pmatrix}^{-1}\begin{pmatrix}
        \frac{w_dz_d}{\gcd(w_d,z_d)} & 0\\
        0 & \frac{x_d y_d}{\gcd(x_d,y_d)}
    \end{pmatrix}
\end{equation}
which now gives
\begin{equation}\label{SL2Q interface}
   \begin{split}
       S_{3d}&=-\frac{i}{2\pi}\int \begin{pmatrix}
        dA & dA'
    \end{pmatrix}\begin{pmatrix}
        \frac{w_d z_d}{\gcd(w_d, z_d)}& 0\\
        \frac{-w_d z_n}{\gcd(w_d,z_d)} & \frac{x_d y_n}{\gcd(x_d,y_d)}
    \end{pmatrix}\begin{pmatrix}
        \Phi\\ \Psi
    \end{pmatrix}\\
    &+\frac{1}{2} \begin{pmatrix}
        \Phi & \Psi
    \end{pmatrix}\begin{pmatrix}
      \frac{w_n z_n w_d z_d}{\gcd(w_d,z_d)^2}  & \frac{-w_n y_n z_d x_d}{\gcd(w_d,z_d)\gcd(x_d,y_d)}\\
       \frac{-w_n y_n z_d x_d}{\gcd(w_d,z_d)\gcd(x_d,y_d)} & \frac{x_n y_n x_d y_d}{\gcd(x_d,y_d)^2}
    \end{pmatrix}d\begin{pmatrix}
        \Phi \\ \Psi
    \end{pmatrix}\,.
   \end{split}
\end{equation}
Also in this case, by fixing 
\begin{equation}
        \begin{pmatrix}
        x & y\\
        w & z
    \end{pmatrix} =
        \begin{pmatrix}
        \frac{x_n}{x_d} & 0\\
        0 & \frac{z_n}{z_d}
    \end{pmatrix} \equiv -\begin{pmatrix}
        \frac{\cN}{\cM} & 0\\
        0 & \frac{\cM}{\cN}
    \end{pmatrix}
\end{equation}
we get
\begin{equation}
    S_{3d}= -\frac{i}{2\pi}\int \Phi \wedge (\cN dA + \cM dA')
\end{equation}
in line with the results in \eqref{eq: rascaling interface}.

Let us finally emphasize that for a fixed $\begin{pmatrix}
        x & y\\
        w & z
    \end{pmatrix} \in \text{SL}(2,\bQ)$ we can have different inequivalent choices of the matrix $\mathcal{D}$. Indeed, the argument of gauge invariance is not enough to uniquely fix the interface between $\mathcal{B}_{4d}$ and $\mathcal{B}_{4d'}$.
    As an example, while still keeping the same form as in \eqref{eq: D sl2q}, one can obtain a different $\mathcal{D}$ by sending $z_{n,d} \to \kappa z_{n,d}$  with $\kappa \in \bZ$. This redundancy is related to the fact that one can stack the interface with condensation defects without altering the $SL(2,\bQ)$ transformation implemented by the interface. In particular, if $\begin{pmatrix}
        x & y\\
        w & z
    \end{pmatrix} = \unit$, the interface obtained is
    \begin{equation}
        S_{3d}=-\frac{i}{2\pi}\int \kappa(A-A')\wedge d\Phi\, 
    \end{equation}
    which, as seen in section \ref{sec: condensation defect}, corresponds to the condensation defect obtained by gauging a $\bZ_{\kappa}$ subgroup of the electric symmetry. If we want minimal interfaces, i.e. without condensation defects, we need to consider irreducible fractions in the parametrization \eqref{SL2Q parametrization}.

\subsection{3d physical boundary and slab compactification}\label{relation to kapustin}
The last ingredient is the 3d physical corner. To reproduce a generic conformal boundary condition, as introduced in Sec. \ref{interacting bc}, we need to add the 3d action
\begin{equation}
    S_{3d\ phys}=S_{CFT} + s' \int i A \wedge * J_{3d}\, ,
\end{equation}
with $s' \in \bZ$. We notice that in this setup there is no need to add topological couplings, as the topological side of the SymTFT generates them automatically.
After a slab compactification, the edge mode $s' A$ can be interpreted as a dynamical gauge field for the $U(1)$ symmetry of the boundary theory $u^T \Phi$. 
 
We can now check that our SymTFT setup correctly reproduces the generic set of boundary conditions introduced in this work. In other words, we wish to make explicit the relationship between the parameters $k,v,p$ and $x,y,w,z$. The reason for doing so is not only to highlight that our SymTFT setup is correct as one would expect, but also to demonstrate the local and global properties of the 3d boundary condition in terms of the parameters $x,y,w,z$.

After the slab compactification, the two 3d corners combine to give the boundary action 
\begin{equation}
    S_{3d} = S_{CFT}+ u^T \int i \overline{\Phi} \wedge * J_{3d}+\frac{i}{2\pi} \int v^T A' \wedge d\overline{\Phi} + \frac{k}{2}  \overline{\Phi }^T \wedge d \overline{\Phi}
\end{equation}
where 
\begin{equation}
    \begin{split}
        &\overline{\Phi}=\begin{pmatrix}
        A\\ \Phi \\ \Psi
    \end{pmatrix}\, ,\qquad u=\begin{pmatrix}
        s'\\ 0 \\ 0
    \end{pmatrix} \, , \qquad v=\begin{pmatrix}
        0\\
        \frac{-w_d z_n}{\gcd(w_d,z_d)} \\
        \frac{y_n x_d}{\gcd(x_d,y_d)}
    \end{pmatrix}\, ,\\ &k=\begin{pmatrix}
        0 & \frac{w_d z_d}{\gcd(w_d,z_d)} & 0\\
        \frac{w_d z_d}{\gcd(w_d,z_d)} & \frac{w_n z_n w_d z_d}{\gcd(w_d,z_d)^2}  & -\frac{w_n y_n z_d x_d}{\gcd(w_d,z_d)\gcd(x_d,y_d)} \\
        0 & -\frac{w_n y_n z_d x_d}{\gcd(w_d,z_d)\gcd(x_d,y_d)}& \frac{x_n y_n x_d y_d}{\gcd(x_d,y_d)^2}
    \end{pmatrix}\,.
    \end{split}
\end{equation}
In the case $s'=0$ we recover the free boundary conditions. We get
\begin{equation}
    \begin{split}
       k^{-1}=\begin{pmatrix}
        \frac{-w_n x_d}{w_d x_n} & \frac{\gcd(w_d,z_d)}{w_d z_d} & \frac{w_n \gcd(x_d,y_d)}{w_d x_n y_d}\\
        \frac{\gcd(w_d,z_d)}{w_d z_d} & 0  & 0\\
        \frac{w_n \gcd(x_d,y_d)}{w_d x_n y_d} & 0 & \frac{\gcd(x_d,y_d)^2}{x_n y_n x_d y_d}
    \end{pmatrix}\,,\quad  k^{-1}v=\begin{pmatrix}
        \frac{-x_d}{x_n}\\ 0\\ \frac{\gcd(x_d,y_d)}{x_n y_d}
    \end{pmatrix}\, , \quad v^Tk^{-1}v=\frac{y_n x_d}{y_d x_n}\,,
    \end{split}
\end{equation}
such that
\begin{equation}\label{final parameters SL2Q}
   P=\frac{x_n y_d}{\gcd(x_n y_d,x_d y_n)}\, ,\quad Q=\frac{x_d y_n}{\gcd(x_n y_d,x_d y_n)}\,, \quad \tilde{r}=\frac{\gcd(x_n y_d,x_d y_n)}{\gcd(x_d,y_d)}\,.
\end{equation}
Consequently, we are able to generate the most general set of $P, Q, \tilde r$ after slab compactification.\footnote{We point out that in the case $y_n=0$, the $k$ used in \eqref{final parameters SL2Q} is degenerate, and instead we should drop out its dependence on $\Psi$. Doing this, we find $P=1,Q=0$ and $\tilde{r}=\frac{z_d}{gcd(z_n,z_d)}$, corresponding to a Neumann boundary condition with a generic non-trivial $\tilde r$. Note that if we send $z_{n,d}\to \kappa z_{n,d}$ this does not change $\tilde{r}$; this is compatible with the fact that this rescaling corresponds to a $\bZ_\kappa$ condensation defect for the electric symmetry. However, since we are dealing with Neumann boundary conditions the above-mentioned gauging is trivial on the boundary.}

The parameters $P$ and $Q$ obtained here only depend on the choice of the $SL(2,\bQ)$ transformation considered and are independent of the choice of matrix $\mathcal{D}$ defining the interface. On the other hand, $\tilde r$ depends on the choice of $\mathcal{D}$. As discussed at the end of Section \ref{sec: B3d topological interface}, keeping the expression \eqref{eq: D sl2q} but considering reducible fractions in the parametrization \eqref{SL2Q parametrization} can lead to insertions of non-trivial condensations defects. These condensation defects can increase the value of $\tilde r$. The minimal value of $\tilde r$ is thus obtained by considering $\gcd(x_n,x_d)=1$ and $\gcd(y_n,y_d)=1$. In this case, the expression for $\tilde r$ can be further simplified to
\begin{equation}
   \tilde{r}=\gcd(x_n,y_n)\,.
\end{equation}
Any other choice of $\mathcal{D}$, not necessarily of the form \eqref{eq: D sl2q}, will lead to a $\tilde r$ proportional to this minimal value. In particular, if $y\neq 0$, it is possible to span all possible values of $\tilde r$ accessible for a fixed $SL(2,\bQ)$ transformation by sending $x_{n,d}\to \kappa x_{n,d}$, $y_{n,d}\to \kappa y_{n,d}$ with $\kappa \in \bZ$.

We conclude by showing that this result is consistent with the one found in Section \ref{sec: 3}. To do that it is useful to notice that any $SL(2,\bQ)$ transformation can be decomposed as a sequence of $SL(2,\bZ)$ transformations and a rescaling as follows:
\begin{equation}\label{eq: sl2q decomposition}
    \begin{pmatrix}
        x & y\\
        w & z
    \end{pmatrix}=\begin{pmatrix}
        P & Q\\
        M & N
    \end{pmatrix}\begin{pmatrix}
        \frac{\mathcal{N}}{\mathcal{M}} & 0\\
        0 & \frac{\mathcal{M}}{\mathcal{N}}
    \end{pmatrix}\begin{pmatrix}
        P' & Q'\\
        M' & N'
    \end{pmatrix},
\end{equation}
where
\begin{equation}
    \begin{pmatrix}
        P & Q\\
        M & N
    \end{pmatrix},\begin{pmatrix}
        P' & Q'\\
        M' & N'
    \end{pmatrix}\in \text{SL}(2,\bZ), \quad \cN,\cM \in \bZ\,.
\end{equation}
With this parametrization, the $\tilde r$ in equation \eqref{final parameters SL2Q} becomes
\begin{equation}\label{final tilde r SL2Q}
\begin{split}
    \tilde r=\frac{\gcd(\mathcal{N}^2,Q)\gcd(\mathcal{M}^2,P)}{\gcd(\mathcal{N},Q)\gcd(\mathcal{M},P)}\,.
\end{split}
\end{equation}
The same result can be obtained by starting with a 3d boundary described by Neumann boundary conditions, i.e.\ $Q = 0$, $P = 1$, and $\tilde r = 1$, and then acting with the sequence of transformations in \eqref{eq: sl2q decomposition}. As shown in Section~\ref{sec: 3}, acting with an $SL(2,\mathbb{Z})$ transformation on the Neumann boundary maps it to a new boundary condition characterized by parameters $P$ and $Q$, while keeping $\tilde r = 1$. We then apply a rescaling, which modifies $\tilde r$ according to \eqref{tilde r rescaling}. Finally, the last $SL(2,\mathbb{Z})$ transformation reshuffles the local boundary condition but leaves $\tilde r$ unchanged. In the end, we find that \eqref{final tilde r SL2Q} precisely reproduces the value of $\tilde r$ obtained in this way. This demonstrates that the SymTFT framework provides a more compact and concise description of the data $(P, Q, \tilde r)$ produced after a generic $SL(2,\mathbb{Q})$ transformation.

When $s' \not=0$ we then obtain a conformal boundary condition. In this case, we see that by setting $s' \in \tilde r \bZ$ we get $r/\gcd(sg,r)=1$ so that, as discussed in Section \ref{interacting bc}, all the CFT charged operators are coupled to the bulk.

\section{Discussion on Non-Compact Edge Modes}\label{sec: non compact b.c.}
In this work we have investigated the effect of coupling topological degrees of freedom to the boundary of a single free photon. While so far we have focused on 3d topological theories with $U(1)$ gauge fields, it is natural to extend the discussion to boundary actions that involve both $N$ $U(1)$-valued edge modes $\Phi$ and $n$ $\bR$-valued edge modes $\phi$. In this Section we would like to highlight the main differences of this more exotic case. 

The new boundary action takes the form
\begin{equation}\label{eq: R bdy}
S_{3d}= \frac{i}{2\pi}\int A\wedge (v_U^T d\Phi+v_R^T d\phi)+\frac{1}{2}\Phi \wedge(k_U d\Phi)+\frac{1}{2}\phi \wedge(k_R d\phi)+\Phi \wedge (k_{UR}d \phi)\,,
\end{equation}
where the term $A\wedge dA$ has already been eliminated in favor of an additional $U(1)$ edge mode.
Here $v_R$ is a vector of real numbers, and $k_{UR}$ and $k_R$ are real matrices. In contrast, $v_U$ and $k_U$ have integer entries to ensure invariance under large gauge transformations of the $U(1)$ fields.

As in the purely $U(1)$ case, we have the freedom to redefine the edge modes. More precisely, the transformation
\begin{equation}
\begin{pmatrix}
\Phi\\
\phi
\end{pmatrix}\to \begin{pmatrix}
S & M\\
0 & m
\end{pmatrix}\begin{pmatrix}
\Phi\\
\phi
\end{pmatrix}
\end{equation}
with $S\in SL(N,\bZ)$, $m\in GL(n,\bR)$ and $M\in M_{N,n}(\bR)$ leaves the boundary conditions unchanged.

The local boundary condition imposed by \eqref{eq: R bdy} is formally the same as if all $\phi$ were $U(1)$ fields, with
 \begin{equation}
 v=\begin{pmatrix}
    v_U\\
    v_R
\end{pmatrix} \quad \text{and}\quad k=\begin{pmatrix}
    k_U & k_{UR}\\
    k_{UR}^T & k_R
\end{pmatrix}\,.
\end{equation}
However, since $v$ and $k$ may contain irrational entries, the combination $v^Tk^{-1}v$ is not necessarily rational. This leads to the more general boundary condition
\begin{equation} \label{real fields eom}
\left(\frac{i}{e^2} *F - \frac{\theta}{2\pi} F\right) + X F = 0 \quad,\quad X := v^T k^{-1} v \in \bR\,.
\end{equation}
This condition can also be understood from the perspective of topological interfaces and their action on boundary conditions, as discussed in Section~\ref{sec: 3}. In the presence of non-compact gauge fields, one can consider more general topological manipulations, including the gauging of $U(1)$ symmetries with flat connections. The associated topological interface is then able to map between any two Maxwell theories with arbitrary couplings $\tau$ and $\tau'$ (see \cite{Paznokas:2025epc}), thus enlarging the space of boundary conditions.

The fate of boundary line operators is instead more sensitive to the distinction between compact and non-compact gauge fields. While the sum over fluxes of $\Phi$ imposes non-trivial constraints on the holonomies, the fluxes of $\phi$ are trivial and therefore do not lead to quantization conditions. Moreover, the spectrum of topological and endable lines will be dependent on the value of $X$. There are three separate cases we can consider.
\begin{enumerate}
    \item When $X \notin \bQ$ we see from the equations of motion in \eqref{real fields eom}, using an analysis similar to that used in Section \ref{section 2.1.2}, that the would be topological Wilson and 't Hooft lines are not well-quantized. This means that are no bulk lines that become topological on the boundary, and therefore they do not generate any 1-form symmetry on the boundary. As a consequence no bulk lines can end on this boundary.
    \item When $X=\frac{Q}{P} \in \bQ$ and $v_R\neq k_{UR}^T\, k^{-1}_U\,  v_U$ we find that the lines $(W_Q H_P)^n \, \forall n\in\bZ$ become topological, yet the lack of constraints on the holonomy of the bulk fields implies that the 1-form symmetry on the boundary is a $\bZ$ symmetry.
    \item Finally, when $v_R=k_{UR}^T\, k^{-1}_U\,  v_U$ we can  perform the field redefinition $\Phi\to \Phi-k_U^{-1} k_{UR}\phi$ to rewrite the action in \eqref{eq: R bdy} such that the $\bR$ fields decouple. Thus, the topological and endable operators are those found in Section \ref{section 2.1.2} (with $v=v_U$ and $k=k_U$) and there exists a $\bZ_{\tilde{r}}$ 1-form symmetry on the boundary. 
\end{enumerate}

Boundary conditions of this type can be interpreted as a higher-dimensional generalization of the Friedan boundary conditions for the 2d compact boson \cite{Friedan1999,Gaberdiel:2001zq}. It is well known that Friedan boundary conditions exhibit pathologies, such as a divergent $g$-function and a continuous spectrum of states. Analogous divergences may arise in the present case,\footnote{For 4d BCFTs, one can consider observables analogous to the $g$-function, such as hemisphere partition functions.} due to the presence of non-compact gauge fields. It would be interesting to analyze these pathologies in detail, to determine whether they reveal general features that go beyond their 2d counterparts.

Let us conclude by noting that the use of non-compact edge modes also allows for generalizations to non-Abelian gauge theories. In this case, one may consider the boundary coupling
\begin{equation}
S = \int_{X_4} \frac{1}{4g^2} \mathrm{Tr}(F \wedge * F)
+ \int_{\partial X_4} \frac{i v}{2\pi} \mathrm{Tr}(F \wedge b)
+ \frac{i p}{4\pi} \mathrm{Tr}\left(CS(A)\right)+\frac{ik}{4\pi}\mathrm{Tr}\left(b\wedge Db\right),
\end{equation}
where $A$ is a $G$-valued connection, $F = DA$, $b$ is a vector of $\mathbb{R}$-valued gauge fields transforming in the adjoint of $G$, and $CS(A)$ denotes the Chern–Simons differential. While $p$ is a quantized coefficient, $v$ and $k$ can be rescaled by a redefinition of $b$. Thus, if non-zero, one of the two can be set to one without loss of generality. It would be interesting to investigate how such topological couplings affect the boundary conditions of Yang–Mills theories.

\section*{Acknowledgments}

We are grateful to Ondrej Hulik and Valdo Tatitscheff for collaboration at the initial stages of this work, and to Lorenzo Di Pietro and Pierluigi Niro for useful feedback. We further thank Fabio Apruzzi and Luca Martucci for correspondence on their recent preprint \cite{Apruzzi:2025byj}. A.A.~and R.A.~are respectvely a Research Fellow and a Research Director of the F.R.S.-FNRS (Belgium).  The research of A.A., R.A., G.G.~and E.P.~is funded through an ARC advanced project, and further supported by IISN-Belgium (convention 4.4503.15).

\bibliography{bib}
\bibliographystyle{ytphys}

\end{document}